\documentclass[article]{jHEP3}

\usepackage{epsfig,multicol,bbm}
\usepackage{amsmath}
\usepackage{subfigure}
\newcommand\fverb{\setbox\pippobox=\hbox\bgroup\verb}
\newcommand\fverbdo{\egroup\medskip\noindent%
                      \fbox{\unhbox\pippobox}\ }

\newcommand\fverbit{\egroup\item[\fbox{\unhbox\pippobox}]}

\newcommand{\bs}[1]{\boldsymbol{{#1}}}

\newbox\pippobox

\title{Automation of the matrix element reweighting method}

\author{Pierre Artoisenet$~^a$, Vincent Lema\^itre$~^b$, Fabio Maltoni$~^b$, Olivier Mattelaer$~^{b,c}$\\
$^a$ Physics Department, The Ohio State University, \\
Columbus, Ohio 43210, USA \\
$^b$ Centre for Cosmology, Particle Physics and Phenomenology (CP3) \\
Universit\'{e} Catholique de Louvain\\
Chemin du Cyclotron 2, B-1348 Louvain-la-Neuve, Belgium\\
$^c$ Istituto Nazionale di Fisica Nucleare (INFN),\\
Sezione di Roma Tre, and Dipartimento di Fisica ``Edoardo Amaldi'',\\
Universit\`a degli Studi Roma Tre, I-00146, Roma, Italy\\
E-mails: \email{partois@pacific.mps.ohio-state.edu,
fabio.maltoni@uclouvain.be,
vincent.lemaitre@uclouvain.be,
olivier.mattelaer@uclouvain.be}
}

\preprint{CP3-10-27\\RM3-TH/10-17}

\abstract{
Matrix element reweighting is a powerful experimental 
technique widely employed to maximize the amount of information that can be extracted from a 
collider data set.  We present a procedure that allows to automatically evaluate the weights 
for any process of interest in the standard model and beyond. Given the initial, 
intermediate and final state particles, and the transfer functions for the final 
physics objects, such as leptons,
jets, missing transverse energy, our algorithm creates a 
phase-space mapping designed to efficiently perform the integration of
the squared matrix element and the transfer functions. The implementation 
builds up on MadGraph, it is completely automatized and publicly available. 
A few sample applications are presented that show the capabilities of the 
code and illustrate the possibilities for new studies that such an approach 
opens up.}

\begin{document}

\newpage

\section{Introduction}

Along with the ongoing experimental activity in Run II at the Tevatron 
and at the now operational Large Hadron Collider at CERN, 
in the last years a significant effort by the high-energy community (including both 
theorists and experimentalists) has been devoted to devise new and more efficient strategies 
to identify physics beyond the standard model in collider data.
In many of the new physics scenarios, new states exist at TeV scale
that decay very quickly, and are not expected to leave any trace that can be 
reconstructed in the detector.  Hence their existence and properties 
must be inferred from the distributions and properties of standard model
particles that can originate from the decay of heavier not-yet-discovered resonances.

The problem of identifying such decay patterns and from those of measuring the properties of the new states
is particularly intricate when the expected experimental signatures involve a complex final state, typically with several jets, leptons and missing energy. The latter, in particular, characterizes 
many of the models that aim at providing a  candidate for dark matter consistent with the 
present observations.  Several methods have been developed during the last few years
to improve on and eventually overcome this difficulty.
For the purpose of identifying new physics, it is quite natural to first consider an 
approach that is not biased by strong theoretical assumptions,
as the current knowledge of the viable theories that could lead to
the production of new particles, is somehow limited.
In this context, different methods have been proposed
to measure the mass spectrum  of the new states in a model-independent way: 
specific observables  are suggested  and built
that are mostly sensitive to the masses of the new heavy resonances entering 
the decay chains. The final power of a given method is a balance between 
how well the information of the visible quantities is exploited to constrain 
the unknown  masses and the dependence on the experimental and theoretical systematic uncertainties.
Examples in the literature include the  end-point method
 ---that is based on the 
end-point regions of the invariant mass distributions built from
visible particles---, and the polynomial method
---that attempts to reconstruct the whole event from 
the visible momenta---. These two methods can also be  combined to give a better 
constraint on the mass spectrum. The $M_{T2}$ method~\cite{Lester:1999tx,Barr:2003rg}, and
its generalizations/recent developments
---though based on
more complicated observables--  also follow the same philosophy.
For a recent review on these kinematics methods see~\cite{Barr:2010zj} 
and references therein.

These model-independent methods will be determinant 
in constraining the mass spectrum of new resonances. However, {\it by construction}, 
most of them  will not exploit or provide any information on other properties of  the new particles,
such as spin and coupling structure. As another example, the precise measurement of the absolute mass 
of each particle entering a specific decay chain that ends with two missing particles remains
challenging, especially in the case of short-length decay chains.  
In this context, it is useful to consider complementary and model-dependent tools 
for the investigation of properties of the new physics states.

Matrix element reweighting~\cite{Kondo:1988yd,Kondo:1991dw,Kondo:1993in,Dalitz:1991wa,Dalitz:1992np,Goldstein:1992xp,Dalitz:1992bx} is an 
example of such a method. It dramatically differs from the previous ones in that
it requires  at least one theoretical assumption as a starting point. 
Each assumption specifies the rules 
needed to compute the probability distribution associated with 
the process under study. Following a Bayesian approach, the method
assigns a probability to each hypothesis given a sample of 
experimental events, and in this way provides a discriminator
among the different hypotheses. Another important feature of the 
method is that it makes maximal use of both experimental information and the 
theoretical model (via the amplitude) on an event-by-event basis. 
This optimal use of the experimental data as well as the 
theoretical knowledge opens the door to new studies not only on mass 
measurements but also on the identification of the spin and coupling type
of new particles.

The matrix element method has been extensively exploited in the last decade.
The best known example is its application to top-quark-pair production 
---investigated both by the CDF and D$\emptyset$ collaborations \cite{Abazov:2004cs,Abazov:2006bd,Abulencia:2007br,Abulencia:2006ry,Abazov:2006bg}---
which has led to the single most precise measurement of the top-quark mass.
Recently, it has contributed to the observation of single-top production~\cite{Aaltonen:2009jj,Abazov:2009ii} and to 
set an upper limit on the boson production of a 
standard model Higgs~\cite{hww:2009pt},
currently excluded in the mass region 158 GeV $<m_H<$ 175 GeV at 95\% confidence level~\cite{:2010ar}.

In principle, the matrix element method is expected to provide a powerful investigation tool 
in many other analyses, in particular those dedicated to the search of new resonances and the study of their properties.
However, in practice, its application is not straightforward. In order to evaluate the weights 
to be attached to each experimental event, a difficult convolution 
of the theoretical information on the hard scattering  ({\it i.e.},  the matrix element squared) with the experimentally available information on the final state
(encoded in the so-called transfer functions) has to be undertaken.
The numerical efficiency (and therefore the speed) of such integration is currently a serious limitation. 
The matrix element  squared as well as  the transfer functions present  variations by 
several orders of magnitude in different regions of the phase space. 
To overcome this difficulty, the integration technique has to be efficiently adapted to the shape of the integrand.
To our knowledge, this problem has only been solved in very specific cases.

In this work, we propose a general algorithm aimed at evaluating the weights 
appearing in the matrix element method. Given an arbitrary decay chain and the associated 
transfer function,  our procedure first  automatically assigns the optimized phase-space 
mappings designed to match as much as possible the peaks in the integrand, and then 
performs the phase-space integrations to evaluate the weights. Our implementation, which is fully automatic, 
is based on MadGraph, as  it uses its matrix element amplitudes and  the information on the topology of the diagrams. We dub the corresponding public code MadWeight.  

The paper is organized as follows. In Section~\ref{sec:def_weight} 
we review the basic features of the  matrix element method. 
In Section~\ref{computation} we expose our algorithm for the computation 
of the weights in the matrix element method. 
We present some illustrations in Section~\ref{applications} 
and our conclusion in the last Section.

\section{The matrix element method}

\label{sec:def_weight}

As mentioned in the introduction, 
the matrix element method is a procedure to extract theoretical information, in the form of  set of parameters $\bs \alpha$  from a sample of experimental events.\footnote{Normally $\bs \alpha$ labels the different parameters in a given model (such as, for example, a mass or the value of a coupling). In this paper, however, we use a more general definition that also includes labelling different physics models.}  Let
us identify an event by the set $\bs x$ of experimentally available quantities (such as transverse momenta, rapidities, and so on). For each observed event a conditional probability $P(\bs x |\bs \alpha)$,{\it i.e.}, a weight, is built that quantifies the ``agreement''  between the theoretical framework  $\bs \alpha$  and the experimental event $\bs x$. In the computation of the weights, one factorizes high-energy effects associated with the production of a parton-level configuration $\bs y$ into a calculable probability $P_\alpha(\bs y)$.
The evolution of the parton-level configuration $\bs y$ into a reconstructed event $\bs x$ in the detector is modeled by a transfer function $W(\bs x, \bs y)$. As a result, the weight of a specific event $\bs x$ is of the form
\begin{equation}
\label{weight_def00}
P(\bs x |\bs \alpha) = \int d{\bs y} P_\alpha(\bs y) W(\bs x,\bs y).
\end{equation}
In the specific case of a hadron collider, the parton-level probability $P_\alpha(\bs y)$ can be expressed as a product of the squared matrix element $|M_\alpha|^2(\bs y)$, the parton distribution functions (pdf's) $f_1(q_1)$ and $f_2(q_2)$ and the phase-space measure $d \Phi(\bs y)$, such that the weight reads 
\begin{equation}
\label{weight_def}
P(\bs x |\bs \alpha)=\frac{1}{\sigma_{ \alpha}} \int d \Phi(\bs y)dq_1 dq_2 f_1(q_1) f_2(q_2) |M_{ \alpha}|^2 (\bs y)  W(\bs x,\bs y)\,.
\end{equation} 
The normalization by the total cross section $\sigma_{ \alpha}$  in Eq.~(\ref{weight_def}) ensures that $P(\bs x |\bs \alpha)$ is a probability density\footnote{We assume that the transfer function is also normalized to $1$.}: $ \int P(\bs x | \bs \alpha) d\bs x=1$.
Once this probability density has been computed for each event $\bs x_i$,  the most probable value for $\bs \alpha$ can be obtained through a likelihood maximization method. \\

Eq. (\ref{weight_def}) is central to this paper as it provides an explicit definition of the weight to be associated with a given event  in terms of the convolution of tree-level matrix element, the pdf's and the transfer functions. One of the main working assumptions in the application of the matrix  element method is that the transfer functions 
are "factorisable",{\it  i.e.}, they can be written as the product of single-particle resolution functions
\begin{eqnarray}
W(\bs x,\bs y)&=&\prod_{i=1}^{n}W_i( x^i,y^i),
\label{eq:TF1}
\end{eqnarray}
where $x^i$ and $y^i$ stand for  the measured quantities and the phase-space variables associated with the particle $i$, respectively. In practice, a further simplification is employed, where the transfer function associated
with a single reconstructed object (such as a jet or a lepton) is written as a product of resolutions associated with the physical quantities measured in the detector:
\begin{eqnarray}
W_i( x^i,y^i)  &=& W_i^E(x^i,y^i)W_i^\eta(x^i,y^i)W_i^\phi(x^i, y^i),
\label{eq:TF2}
\end{eqnarray}
where $E,\eta$ and $\phi$ are the reconstructed energy, rapidity and azimuthal angle.
in most of the general purpose detectors the direction of a visible particle\footnote{Throughout the paper, 
the expression 
\textit{visible particle} refers to a lepton or a jet of which momentum is reconstructed in the detector.} 
is  well measured, so that the associated transfer function can be modeled by a narrow Gaussian. On the other hand, the resolution in energy strongly depends on the particle's type. For leptons it can be taken as a narrow Gaussian function whereas  for jets 
a more involved parametrization of the resolution function is needed.

In this work we provide a general solution to the problem of performing the integration in Eq.~(\ref{weight_def}) in an efficient way.  
To better grasp the challenge that computing the integral in the numerator of Eq.~(\ref{weight_def}) poses, 
it is useful to consider two limiting cases, where the problem simplifies.

First let us imagine to have an ``ideal'' detector that could measure exactly the energies and momenta of all 
final state particles (including normally invisible ones), {\it i.e.}, $W(\bs x,\bs y)=\delta(\bs x- \bs y)$.\footnote{For jets, the transfer functions also include genuine
QCD effects like showering and hadronization. For the sake of the argument, we consider also "ideal" jets where the identification jet/parton
is perfectly unambiguous.} In this case no integration would be 
necessary and the weight in the numerator would be proportional to the corresponding squared matrix 
element, $|M_{ \alpha}|^2 (\bs x)$.  Nowadays, the determination of 
$|M_{ \alpha}|^2 (\bs x)$ at the tree-level  can be done automatically by several public 
codes and poses no difficulty.  So apart from the normalization, discussed below, the weight calculation would therefore be  trivial.

As a second limiting case, one can also consider an ideal  ``no detector" option, {\it i.e.},  choose 
the transfer function  $W(\bs x,\bs y)=1$.  Then the integration would reduce
to the computation of the total cross section,{\it  i.e.}, that of the denominator of Eq.(\ref{weight_def}) as  $P(\bs x | \bs \alpha) = 1$. 
This problem  is not an easy one on its own: the matrix element has a very complicated peak structure, 
corresponding to the propagators of the Feynman diagrams being large. 
However, by observing
that the leading peaks come from the sum of the squares of each diagram,  together with the fact that it is 
always possible to find a parametrization of the phase space in terms of invariants that maps exactly those in the propagators~\cite{Byckling:1970wn}, 
makes the problem  treatable (see for example Refs.~\cite{Kleiss:1994qy,Maltoni:2002qb} and 
the discussion in the following Section).

For a realistic detector the situation is in between the two above, where some particles are well measured
(charged leptons), other less (jets), and some completely missed (neutrinos). 
In this case the integration becomes extremely difficult as it involves an integrand 
with simultaneous peaks in sets of different variables that it is not possible, even in principle, to disentangle.

\section{Computation of the weights}

\label{computation}

The evaluation of multi-dimensional integrals is often approached by standard adaptive 
Monte Carlo techniques. 
These techniques are well illustrated in the computation of total cross sections:
phase-space mappings that ``flatten'' specific peaks in the integrand are 
combined together in a multichannel integration. Here we also follow this approach.
In our case, the phase-space mappings optimized for 
the computation of the weights in Eq.~(\ref{weight_def})
are rather involved because of the complex structure of peaks in the integrand,
as it was discussed in the previous section.

In this section we present our integration procedure and its implementation in a fully general algorithm.
We first recall the basic principle of an adaptive Monte Carlo integration
in Section~\ref{sec:adaptative_int} and
then we describe the phase-space mappings optimized for the computation of the weights in Section~\ref{sec:PS_int}.
We explain how we build a phase-space generator based on these new phase-space mappings in Section~\ref{sec:generator}
and how we combine different phase-space mappings in a multi-channel 
integration in Section~\ref{sec:multi_channel}.  We validate our phase-space generator with several checks 
 in Section~\ref{sec:checks}.

\subsection{Adaptive Monte Carlo techniques}

\label{sec:adaptative_int}

Adaptive Monte Carlo integration
is a powerful numerical technique for the integration
of a highly non-uniform function.
It consists of sampling randomly the volume of integration
according to a probability density that is adjusted
iteratively to the shape of the integrand.
The probability density is parametrized by a separable function
\begin{equation}
p(\bs{z})=p_1(z^1)\, p_2(z^2)\,\dots\, p_d(z^d)
\label{vegas_grid}
\end{equation}
where each factor $p_i$ is a step function.
If such a parametrization of the probability density function
is appropriate to approximate the shape of the
integrand, the adaptive integration procedure
speeds up the convergence by
increasing the density of evaluations in the regions where
the integrand is large.
In the case of a very sharp integrand,
this condition is essentially fulfilled
provided that the strength of each
narrow peak in the integrand is associated with a single
variable that in turn can be mapped onto one variable
 of integration $z^i$. In that case, the integrand expressed in the
parametrization $\bs{z}$ is of the form
\begin{equation}
f(\bs z ) = \left( \prod_{i=1}^{d} f_i(z^i)\right) \times R(\bs z )
\label{Vegas_ideal_case}
\end{equation}
where the functions $f_i$'s may vary abruptly while the ``remainder" non-factorisable function $R(\bs z)$
is essentially flat over the region under integration.

If the integrand expressed in the phase-space mapping $\bs{z}$
presents a structure of sharp peaks that does not follow
the factorized form in Eq.~(\ref{Vegas_ideal_case}),
the adaptive integration procedure is bound  to fail.
However, if enough information about the shape of the integrand is available,
a first change of variables $\bs{z}\rightarrow \bs{z}'=\bs{P}(\bs{z})$ that
rotates the axes of integration can sometimes be applied
 such that in the new phase-space mapping $\bs{z}'$, the importance of each peak
 in the integrand is controlled by a
single variable of integration.
After this change of variables is applied, the integrand expressed in the new variables $\bs{z}'$
is of the form given by Eq.~(\ref{Vegas_ideal_case}),
and the separable density function $p(\bs{z}')$ can be successfully
adapted to the shape of the integrand.

We will use the adaptive Monte Carlo integrator VEGAS~\cite{Lepage:1980dq} to carry out the integration
in Eq.~(\ref{weight_def}). Thus the efficiency in computing the weights will depend
on the parametrization of the phase-space measure that is used in the
adaptive Monte-Carlo integration.
The optimized phase-space mappings are such that for each narrow peak
either in the transfer function or in the matrix element, the variable
that controls the strength of that peak is mapped onto a single variable
of integration in the parametrization of the phase-space measure, in which case the integrand
expressed in that parametrization
has the form given in Eq.~(\ref{Vegas_ideal_case}).

\subsection{The new phase-space mappings}
\label{sec:PS_int}

For the computation of the weights, there is generally
no simple phase-space parametrization that maps all the peaks 
in the integrand and in which the boundaries of the phase-space volume
can be easily expressed. Our strategy is to start from the following standard 
parametrization of the phase-space measure
\begin{equation}
\label{canonicalPS}
d \Phi = \left( \prod_{i=3}^{n} \frac{ |\bs{p}_i|^2 d|\bs{p}_i| \sin \theta_i d \theta_i d\phi_i}{2E_i(2\pi)^3 } \right)  dq_1 dq_2 (2 \pi)^4 \delta^4 \left(p_1+p_2-\sum_{j=3}^n p_j \right)  ,
\end{equation}
where $i=3,\dots n$ labels the final particles.
In this parametrization, the strength of each peak in the transfer function
is already mapped onto a single variable of integration, whereas none of the
propagator enhancement in the squared amplitude is. 
Identifying the Lorentz invariants associated with the Breit-Wigner resonances 
and expressing them as functions of the integration variables 
in Eq.~(\ref{canonicalPS}) is straightforward. The difficult task is then to 
invert these functions in order to derive a phase-space measure that is parametrized
by both these Lorentz invariants and the variables mapping the peaks in the transfer function.
Along with this inversion, the $\delta$ function associated with energy-momentum conservation 
in Eq.~(\ref{canonicalPS}) has to be integrated out.
The resulting phase-space mappings can then be used in an adaptive Monte Carlo integration
to compute the weights. 

These optimized phase-space mappings can be defined by specifying the transformation
of the phase-space measure parametrization in Eq.~(\ref{canonicalPS}) from which they result.
So in this Section, we will describe the expression of this transformation in a generic case,
as it is a convenient way to introduce the new phase-space mappings.
For an arbitrary process, the transformation that leads to the appropriate 
parametrization of the phase-space measure can be carried out by organizing the integration 
variables in the standard parametrization in Eq.~(\ref{canonicalPS}) into different subsets
of variables to which a suitable change of variables is applied. Each subset of variables and 
its associated change of variables will be called a \textit{block} in the following.  

The first phase-space block that needs to be identified is called 
the \textit{main block} (MB), and it includes some of the integration variables 
appearing in Eq.~(\ref{canonicalPS})
to which a transformation is applied so that the $\delta$ function associated 
with energy-momentum conservation is integrated out. The same transformation may also map
some invariants entering in the expression of specific propagators 
to new variables of integration in the expression of the phase-space measure.
The identification of the main block and the form of the associated  transformation of variables
 is discussed in the following Section. The integration variables appearing in Eq.~(\ref{canonicalPS}) that do not belong to the main block also experience a transformation that can be expressed in terms
of \textit{secondary blocks}, as explained in Section~\ref{sec:blob}.

\subsubsection{Identification of the main block}
\label{sec:MB}

The main block includes a certain number of integration variables
among those appearing in Eq.~(\ref{canonicalPS}). These variables are 
adjusted as a function of all other kinematic quantities associated with the decay chain
to enforce the conservation of energy and momentum.

\begin{figure}
\center
\subfigure[]{\includegraphics[width=7cm]{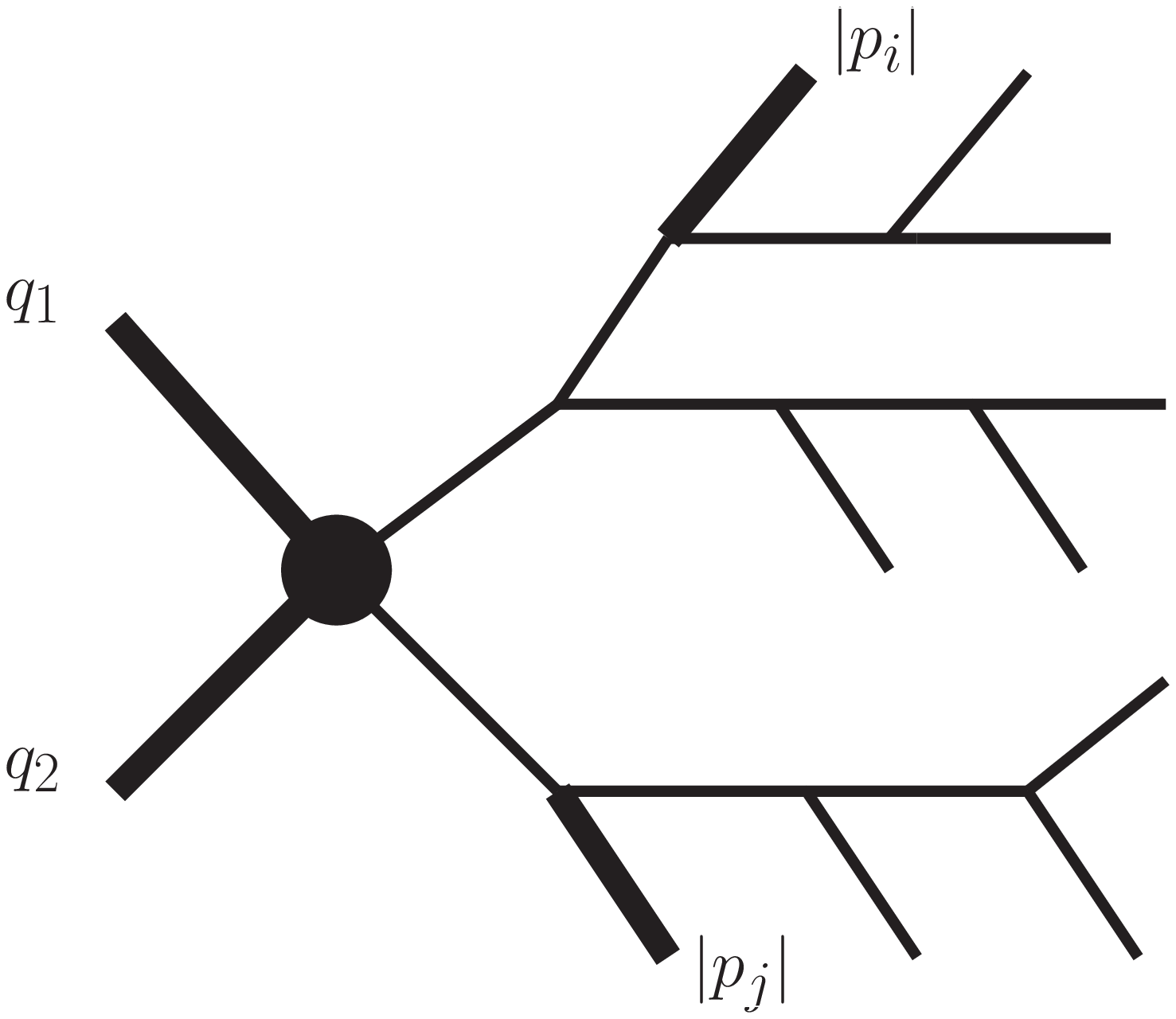}}
\subfigure[]{\includegraphics[width=7cm]{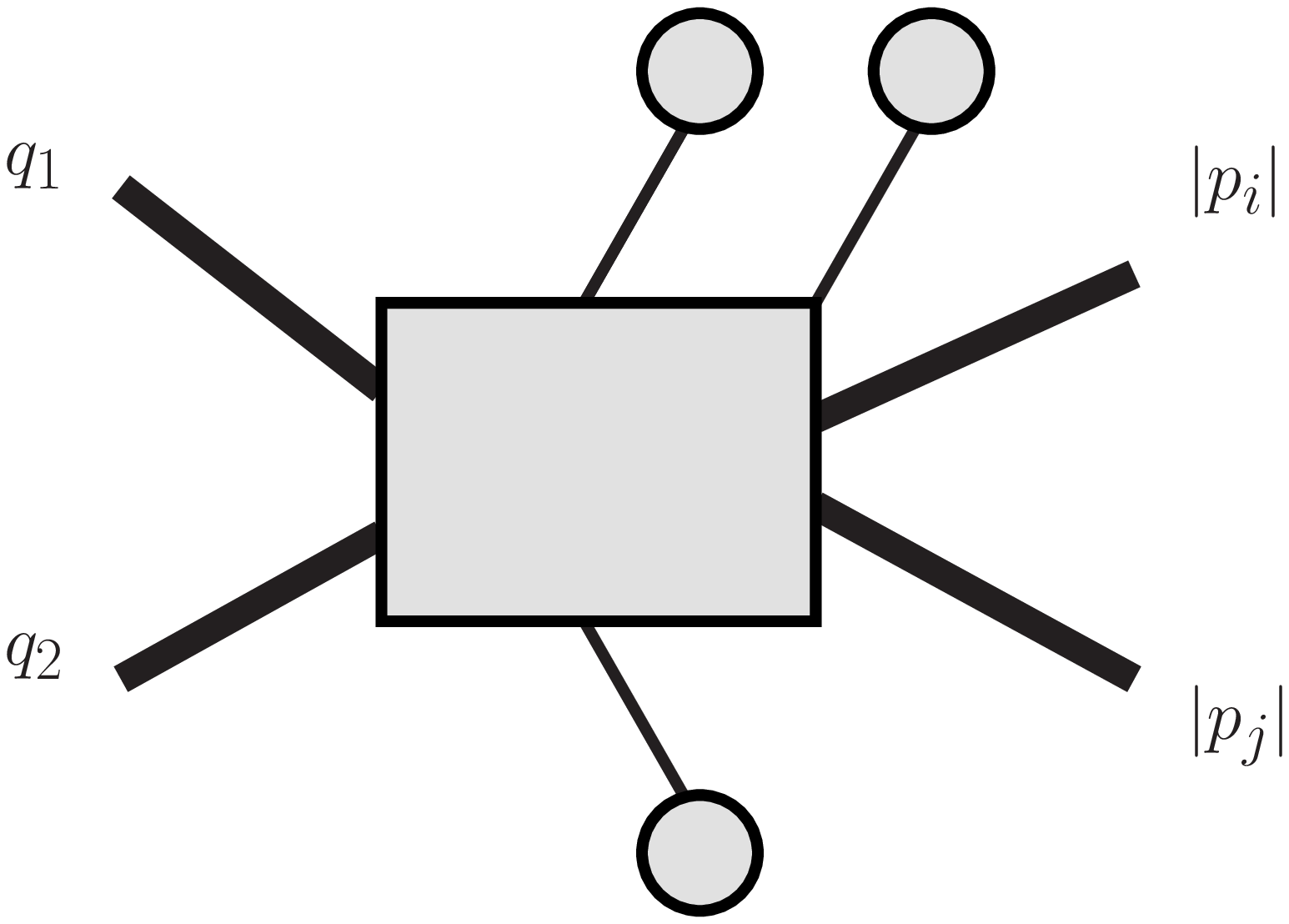}}
\caption{Illustration of a  decay chain with no missing particle: (a) the full topology, (b) the corresponding reduced diagram.
The variables in the main block are written explicitly.}
\label{fig:no_missing_ex}
\end{figure}

We start by discussing the choice of the main block in the case of two specific decay chains, and then generalize to the case of an arbitrary decay chain. We first consider a topology with no missing particle. An example of such a decay chain is illustrated in Figure~\ref{fig:no_missing_ex}(a). In that case, it  is natural to include the initial proton momentum fractions of partons (called Bjorken fractions) in the main block, 
as the integrand does not show any sharp sensitivity in these variables. The angle 
of any visible particle
should be excluded from the main block as it controls the strength of a narrow resolution function $W^\eta$ or $W^\phi$ in Eq.~(\ref{eq:TF2}):
it should therefore be maintained 
as an integration variable to ensure the integration of the associated peak.
 However, normally the resolution in energy is much poorer than the resolution in angles. For this reason, a relatively efficient choice is to complete the main block by adding two momentum variables $|\bs{p}_i|$ and $|\bs{p}_j|$ of particles $i$ and $j$ that are relatively less constrained by the transfer function. In this example, the main block contains exactly four variables, and the effect of the variable transformation is to integrate out the $\delta$ function with these four variables. 
 
\begin{figure}
\center
\subfigure[]{\includegraphics[width=7cm]{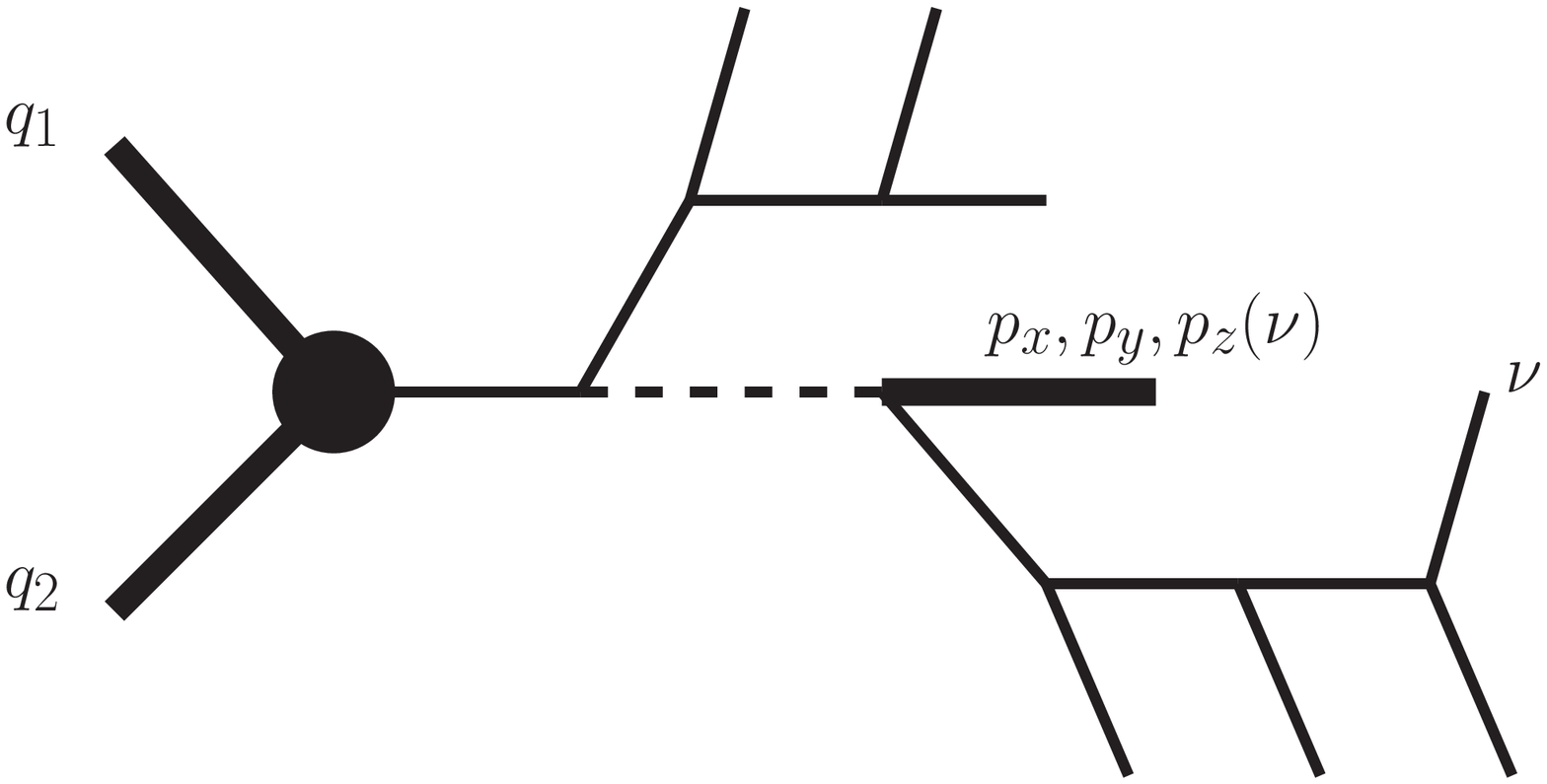}}
\subfigure[]{\includegraphics[width=7cm]{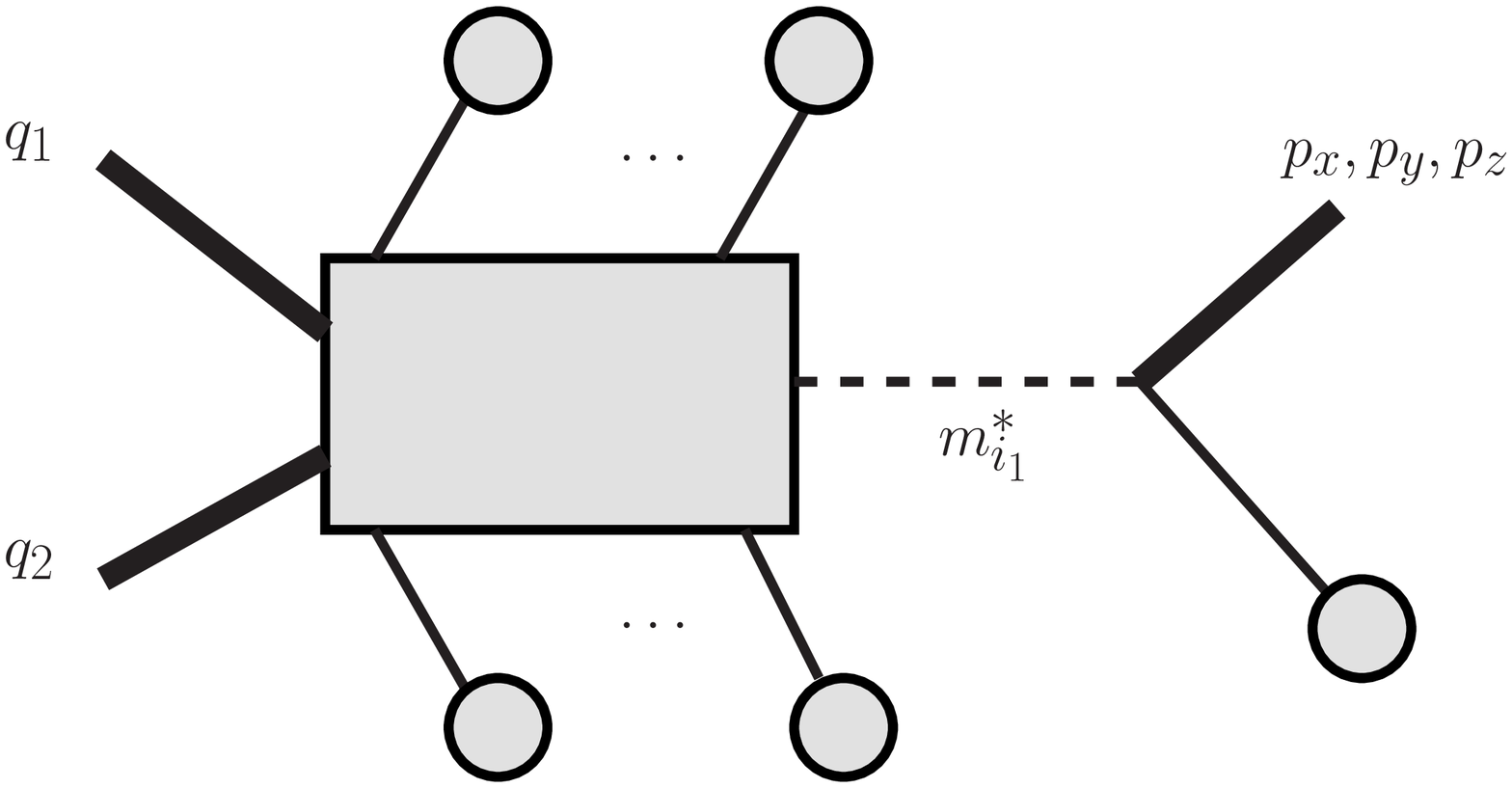}}
\caption{Illustration of a decay chain with two missing particles (identified by the letter $\nu$): (a) the full topology, 
 (b) the corresponding reduced diagram. The initial (resp. final) variables of the transformation 
associated with the main block are written explicitly, and the corresponding legs are shown as thick lines (resp. dashed lines). }
\label{fig:missing_ex}
\end{figure}

We next move on to the case of a decay chain including missing particles in the final state.  The phase-space variables associated with the momenta of missing particles are not directly constrained by the transfer function. Therefore they do not need to be mapped onto variables of integration in the phase-space mapping. We can identify the main block by selecting the momentum components of some missing particles instead of the energies  of visible particles. A specific example of topology with missing particles is displayed in Figure~\ref{fig:missing_ex}(a). 
One way to define the main block is to choose the set including the Bjorken fractions and the momentum components of the missing particle shown as a thick line in Figure~\ref{fig:missing_ex}(a). 
The change of variables associated with this main block remove these five variables from the 
set of integration variables in order to integrate out the $\delta$ function in Eq.~(\ref{canonicalPS})
and to map the invariant mass of the resonance decaying into the missing particle onto a variable 
of integration in the new parametrization of the phase-space measure.

In order to generalize the discussion of the choice of the main block to the case of an arbitrary decay chain, it is useful
to introduce  the following representation of the main block and the corresponding transformation of variables:
\begin{itemize}

\item In a branch of legs with no kinematic variable in the main block,
the decay products of the initial particle in the branch are shrunk into a blob.

\item The variables in the main block are written explicitly and the corresponding legs are shown as thick lines.

\item The new integration variables resulting from the change of variables associated with the 
main block are also written explicitly, and the corresponding intermediate legs are shown as dashed lines.

\item All other intermediate legs that do not touch a blob are hidden behind a rectangular box.

\end{itemize} 
We refer to the resulting graph as the  \textit{reduced diagram}. As an illustration, 
the reduced diagrams for the two topologies shown in Figures~\ref{fig:no_missing_ex}(a) and \ref{fig:missing_ex}(a) are displayed in Figures~\ref{fig:no_missing_ex}(b) and \ref{fig:missing_ex}(b), respectively. In general a blob in a reduced diagram may hide a complicated branch of particles. 
But the change of variables associated with the main block is parametrized only by the
total momentum of each blob, it does not depend on the structure inside the blob.

We use these reduced diagrams to represent the main block in general. 
The minimum number of variables in the main block is four. After the variable transformation associated with the main block is applied, the  $\delta$ function in Eq.~(\ref{canonicalPS}) is integrated out with these four variables that therefore do not appear in the new phase-space mapping resulting from this transformation. The main block may contain $p>4$ integration variables. 
The transformation that is applied in that case removes all these $p$ variables from the set of integration variables appearing in the parametrization of the phase-space measure, and introduces $p-4$ new variables of integration. Each of these new variables map a Lorentz invariant that controls the strength of a specific propagator in the matrix element. Thus the variable transformation associated with the main block not only enforces the conservation of total energy and momentum, but also  may possibly 
optimize the parametrization of the phase-space measure for the integration of some specific Breit-Wigner enhancements. 

As the variables in the main block are removed from the phase-space mapping after the corresponding transformation is applied, 
an integration variable in the standard parametrization that controls the strength of a narrow peak in the transfer function is preferentially not included in the main block, otherwise the phase-space mapping after transformation would loose track of this variable and would be inappropriate for the integration of the corresponding peak. From this observation, it is clear that the choice of the main block will act upon the efficiency 
of the Monte Carlo integration.

\begin{figure}
\center
\subfigure[MB A]{\includegraphics[width=7cm]{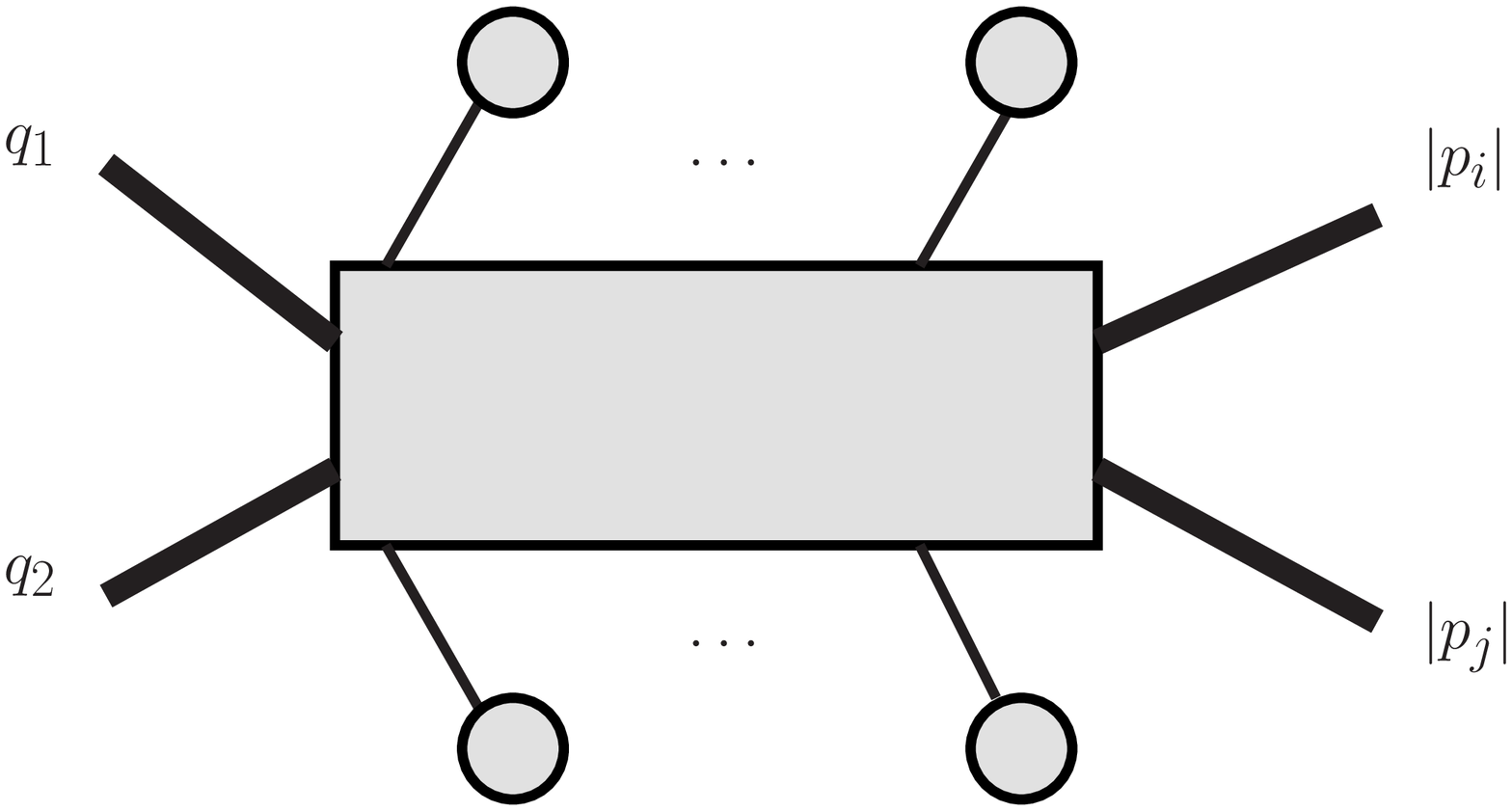}}
\subfigure[MB B]{\includegraphics[width=7cm]{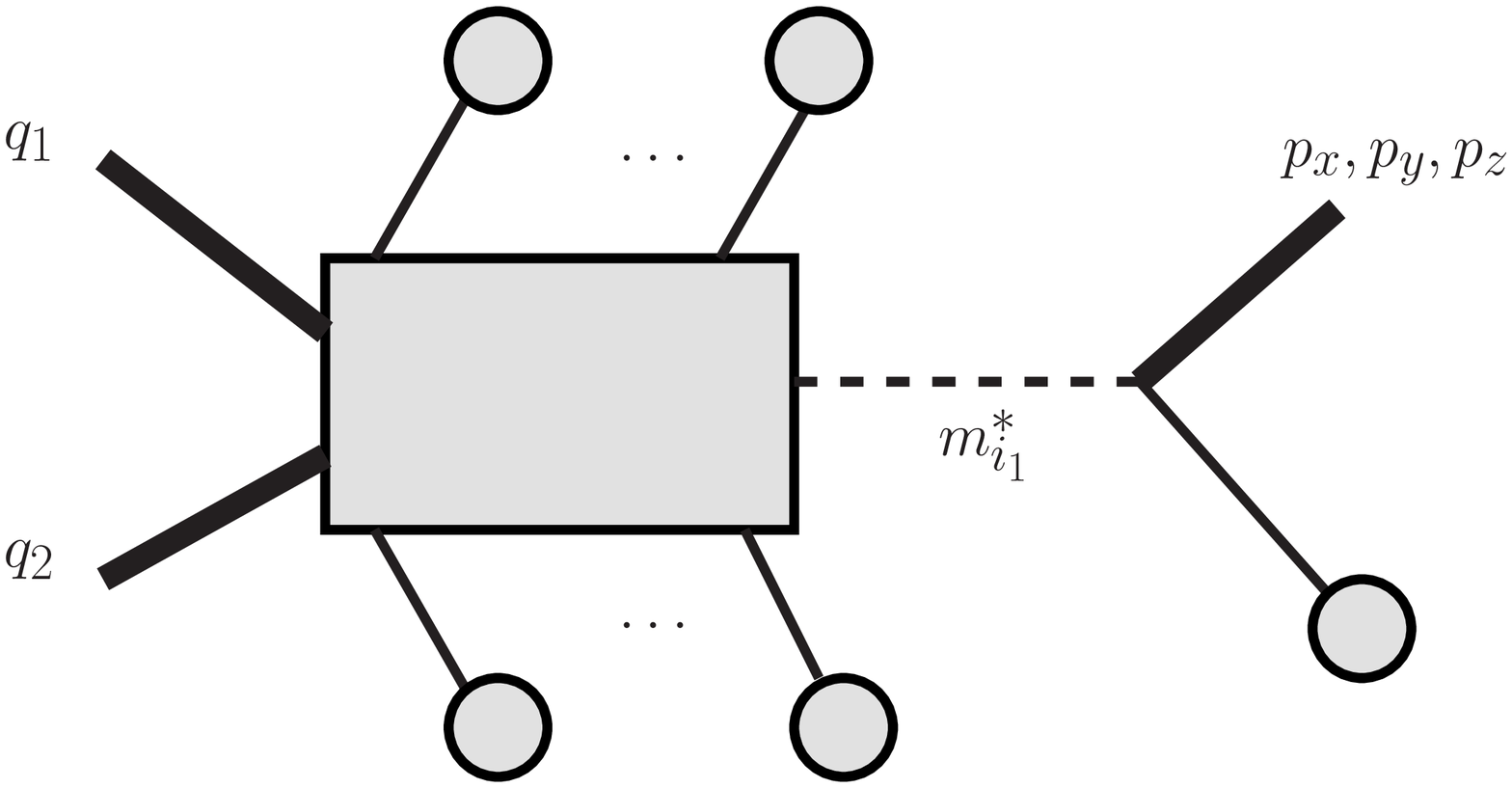}}
\subfigure[MB C]{\includegraphics[width=7cm]{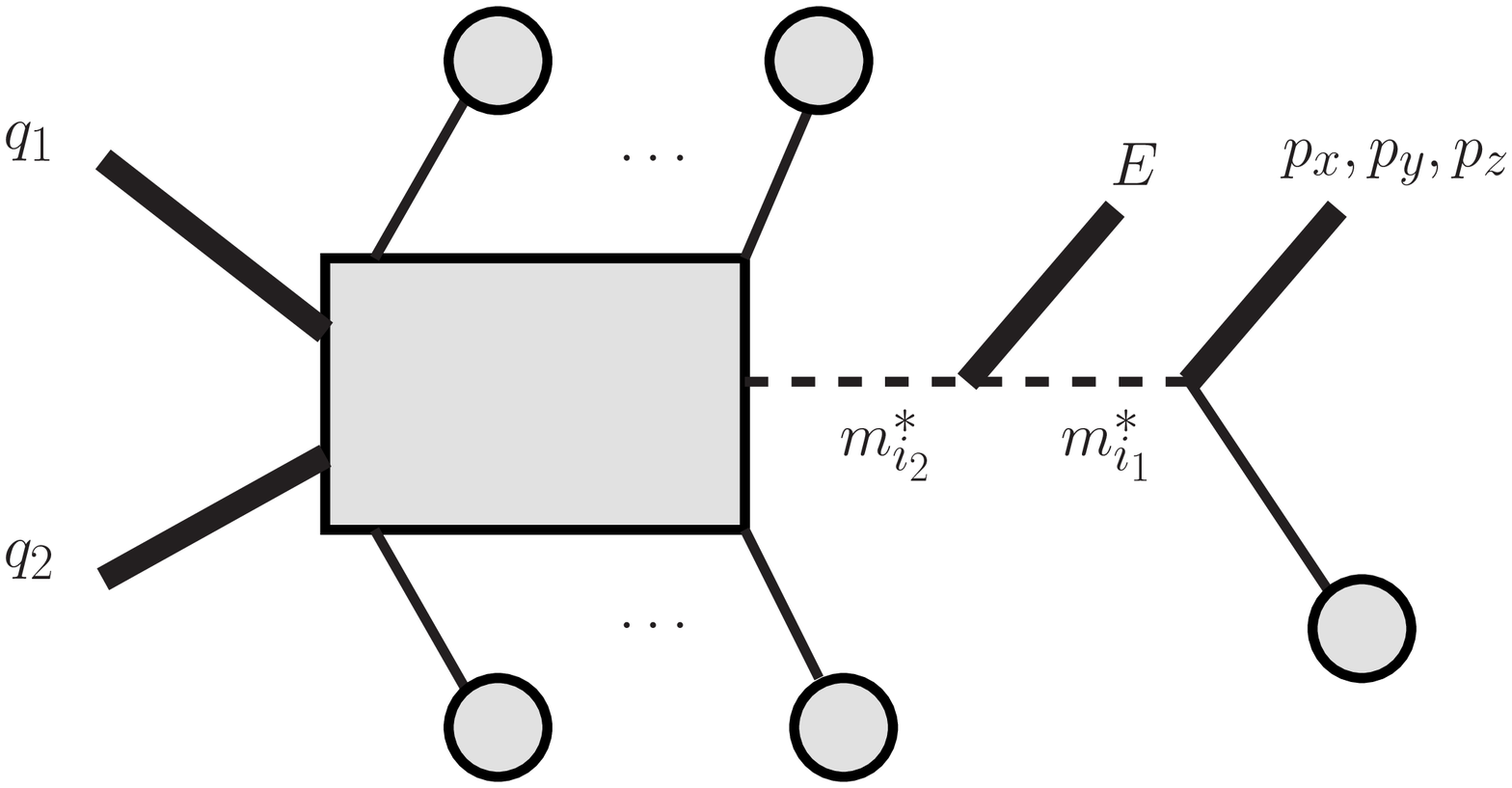}}
\subfigure[MB D]{\includegraphics[width=7cm]{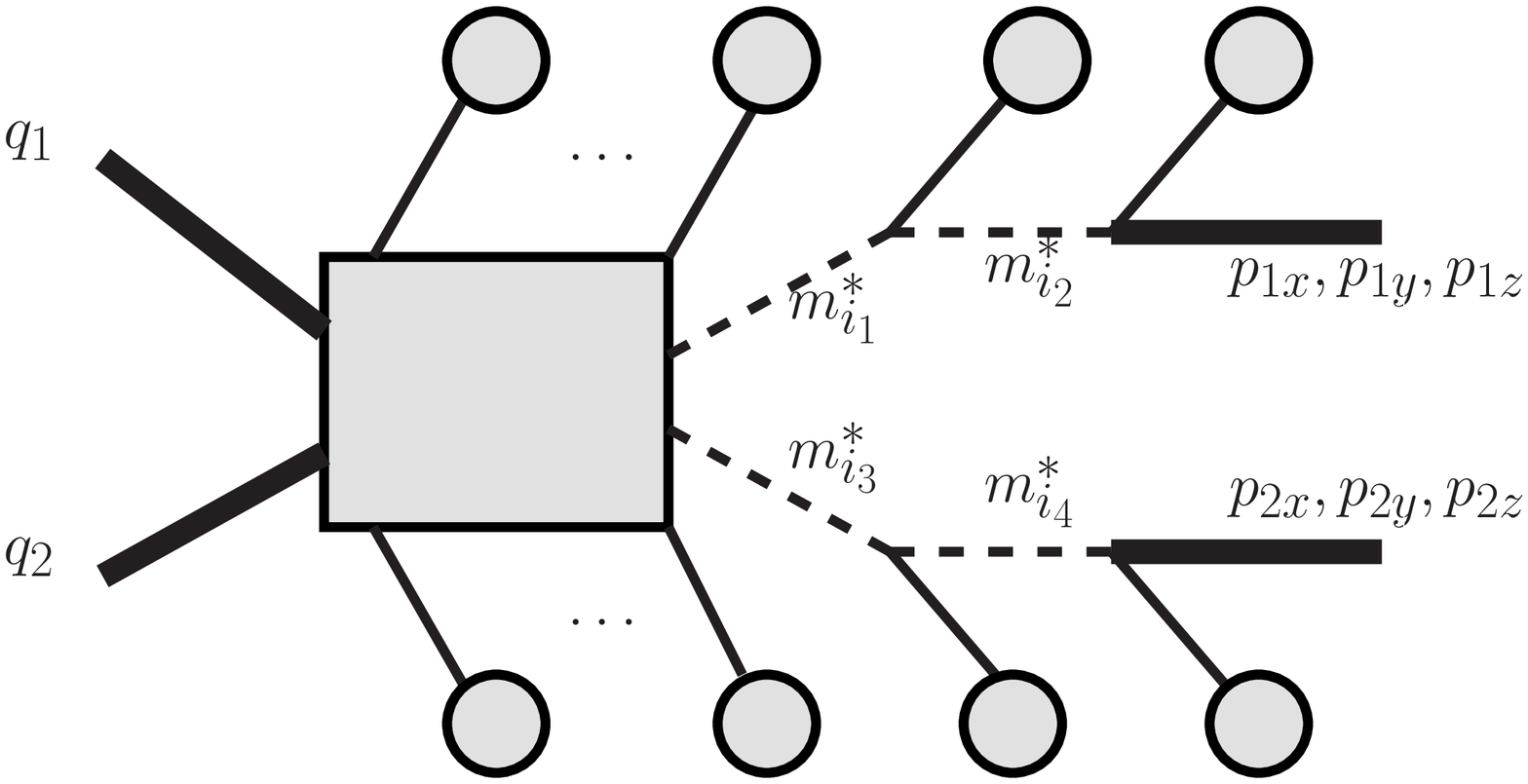}}
\subfigure[MB E]{\includegraphics[width=7cm]{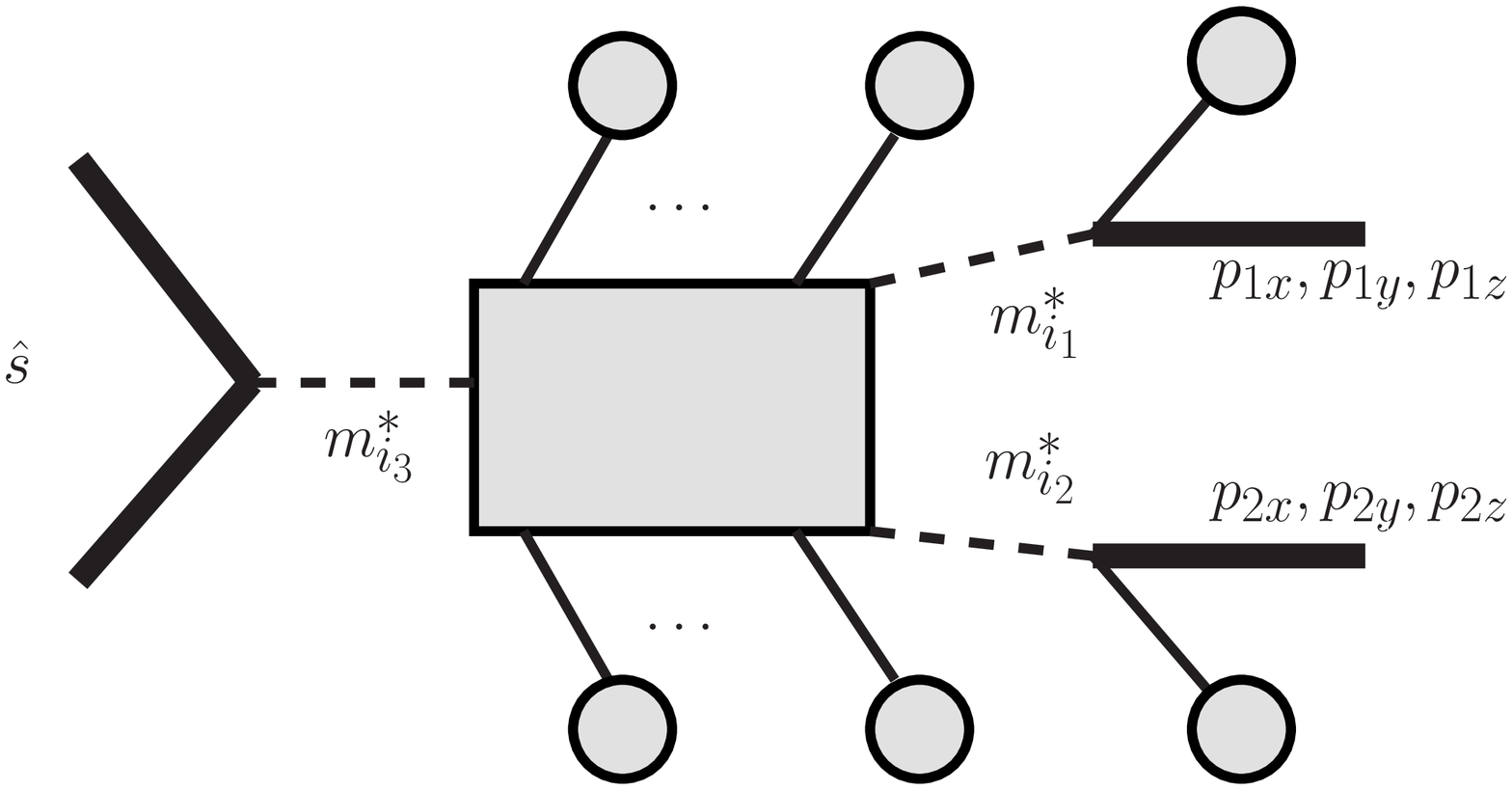}}
\subfigure[MB F]{\includegraphics[width=7cm]{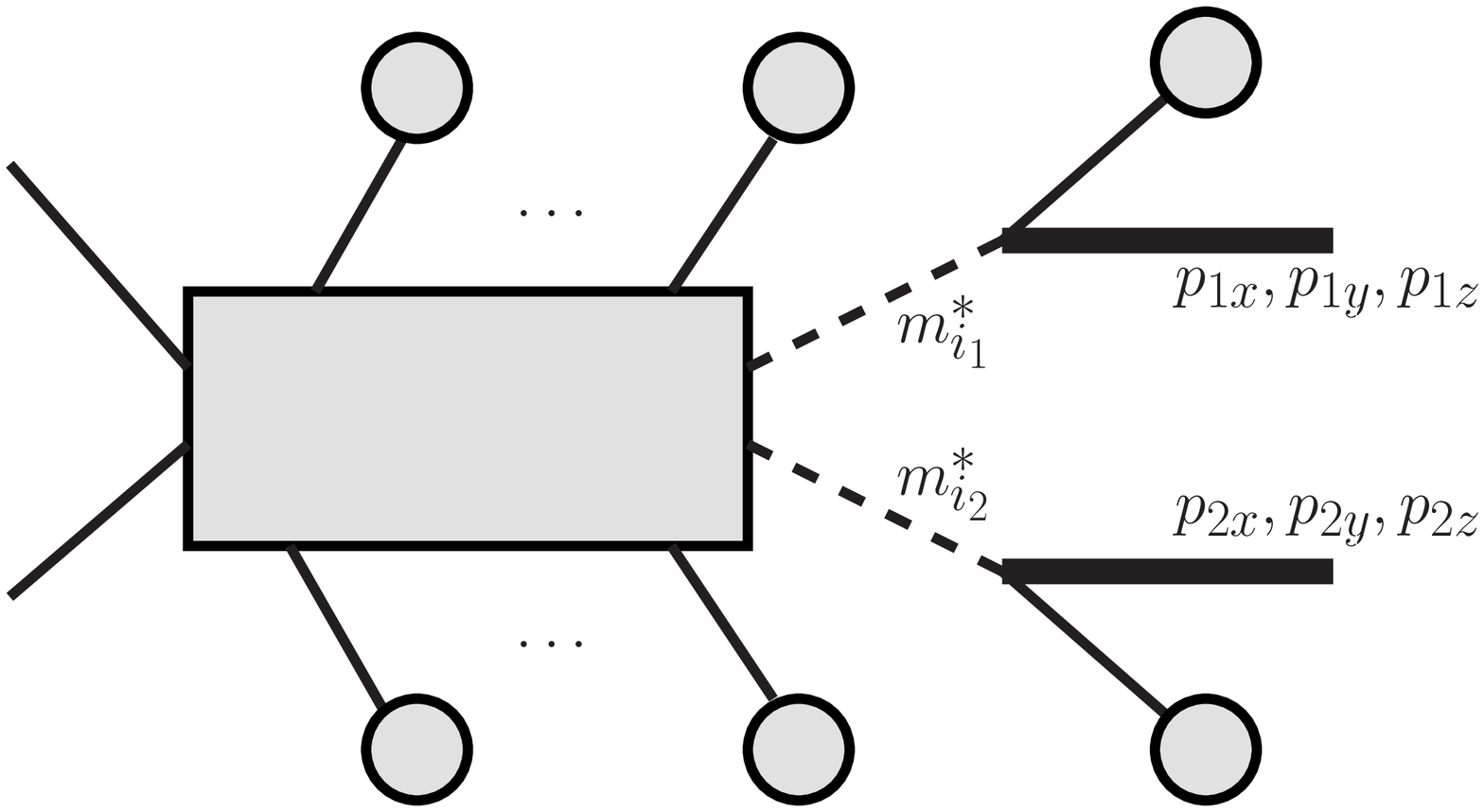}}
\caption{The reduced diagrams representing the six main blocks
that have been investigated in our procedure. \label{fig:main_blocks}}
\end{figure}

Each of the main blocks treated in our code is illustrated
by a reduced diagram in Figure~\ref{fig:main_blocks}.
Their number is restricted because we only keep the main blocks for which 
the corresponding change of variables is invertible analytically.
 The corresponding formulas are discussed in Appendix~\ref{appendix:MB}.

\begin{description}
\item[MB A.]  The transformation removes the Bjorken fractions $q_1$ and $q_2$ and 
the norm of the three-momenta $\bs p_i$, $\bs p_j$ of two visible particles from the
set of integration variables in the parametrization of the phase-space measure. \\
Example: $pp \rightarrow ZZ \rightarrow 4 j$.

\item[MB B.] The transformation removes the Bjorken fractions $q_1$ and $q_2$ and 
the 3-momentum of a missing particle from the
set of integration variables in the parametrization of the phase-space measure.
The new integration variable is the invariant mass of the particle decaying into the missing particle. \\
Example: $pp \rightarrow Z(W^+ \rightarrow l^+\nu) $.

\item[MB C.] 
The transformation removes the Bjorken fractions $q_1$ and $q_2$, the 3-momentum of a missing 
particle and the energy of a  massless visible 
particle\footnote{In the case the particle is massive, the corresponding change of variables turns out to be not analytically invertible.}
from the set of integration variables in the parametrization of the phase-space measure.
The new integration variables are the Lorentz invariants $m_{i_1}^*$ and $m_{i_2}^*$
associated with the mother particles decaying into the missing and the massless particles,
respectively. \\
Example: $pp \rightarrow [t \rightarrow b(W^+ \rightarrow l^+\nu)] [\bar t \rightarrow \bar b(W^- \rightarrow jj)]$ with massless $b$ quarks.

\item[MB D.] 
The transformation removes the Bjorken fractions $q_1$ and $q_2$ and the 3-momenta of two missing 
particles from the set of integration variables in the parametrization of the phase-space measure.
The new integration variables are the Lorentz invariants
 $m_{i_1}^*$, $m_{i_2}^*$, $m_{i_3}^*$ and $m_{i_4}^*$ associated with the first and second
 mother particles of each missing particle. \\
Example: $pp \rightarrow [t \rightarrow b(W^+ \rightarrow l^+\nu)] [\bar t \rightarrow \bar b(W^- \rightarrow l^-\bar \nu)]$.

\item[MB E.] 
The transformation removes  the 3-momenta of two missing 
particles from the set of integration variables in the parametrization of the phase-space measure.
The new integration variables are the Lorentz invariants
 $m_{i_1}^*$ and $m_{i_2}^*$  associated with the mother particles of each missing particle.
The integration over the Bjorken fractions is expressed as an integration over the invariant mass and the rapidity of the 
colliding partons. \\
Example: $pp \rightarrow H \rightarrow (W^+ \rightarrow l^+\nu) (W^- \rightarrow l^-\bar \nu)$.

\item[MB F.] 
The transformation removes  the 3-momenta of two missing 
particles from the set of integration variables in the parametrization of the phase-space measure.
The new integration variables are the Lorentz invariants
 $m_{i_1}^*$ and $m_{i_2}^*$  associated with the mother particles of each missing particle. \\
Example: $pp \rightarrow (W^+ \rightarrow l^+\nu) (W^- \rightarrow l^-\bar \nu)$.
\end{description}

For a given decay chain, the main block can be chosen in several different ways, 
possibly with some more efficient than others.
Roughly speaking a specific main block is appropriate for the computation of the weights
provided that it does not contain a variable that controls the strength
of a very sharp resolution function, and provided that no very sharp 
Breit-Wigner distribution is included in the square of the reduced diagram. 
This second condition comes from the fact that 
 none of the invariants  entering into the expression of the 
propagators  inside the box 
is mapped onto a single variable of integration
 in the new parametrization of the 
phase-space measure, so that parametrization 
is not appropriate for the integration of the corresponding propagator enhancements.

It should be stressed once again that each transformation 
of variables that is applied to the variables in main block 
 has been implemented in the code analytically: for an arbitrary phase-space point, given
the momenta of all the legs ending by the blobs
and the invariant mass of each leg represented by a dashed line
in the reduced diagram, the variables in the main block are
determined by means of analytical expressions (see 
appendix~\ref{appendix:MB}).
We have not explored further the possibility to use  
 numerical procedure for this step. So any change of variables that is not 
invertible analytically 
 has been excluded in our algorithm. 

\begin{figure}
\center
\resizebox{!}{5cm}{ \includegraphics[]{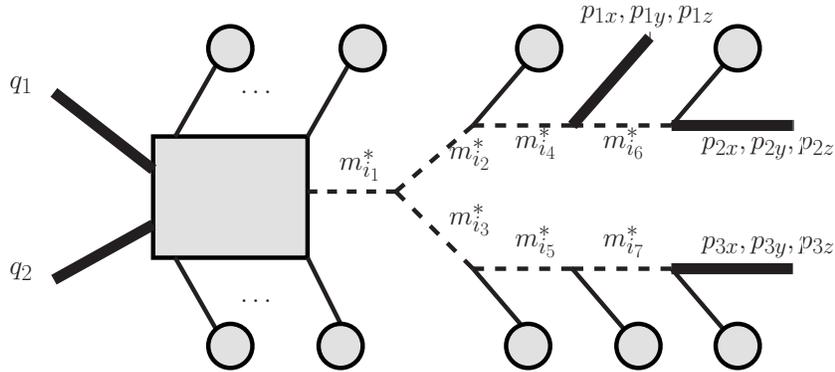}}
\caption{A example of reduced diagram
for which the transformation that is applied to the MB cannot be inverted by means of analytical formulas. \label{fig:non_analytic_block}}
\end{figure}

For example, the main block displayed in Figure \ref{fig:non_analytic_block} with three missing 
particles has not been considered. In principle, the 3-momenta of the three missing particles and the Bjorken fractions could be adjusted to satisfy eleven constraints induced by the seven resonances and the conservation of 4-momentum. As this adjustment cannot be done by means of analytical 
expressions, this case is dealt with the  
main block  B or C in our procedure.

\subsubsection{Identification of the secondary blocks}
\label{sec:blob}

Once the main block has been defined and the corresponding transformation applied,
the parametrization of the phase-space measure associated with the $m$ external legs in all the blobs
of the reduced diagram is still the standard one:
\begin{equation}
\label{eq:canonicalPS_v2}
\prod_{i=1}^{m} \frac{ |\bs{p}_i|^2 d|\bs{p}_i| \sin \theta_i d \theta_i d\phi_i}{2E_i(2\pi)^3 } .
\end{equation}
The parametrization of the phase-space measure is further transformed by
organizing the integration variables in Eq~(\ref{eq:canonicalPS_v2})
into \textit{secondary blocks}, {\it i.e.}, into subsets of 
variables, each of them being subject to a specific change of variables. 
The change of variables of the simplest block is just the identity, in which 
case the variables in this block are maintained in the parametrization of the phase-space measure
in Eq.~(\ref{eq:canonicalPS_v2}). 
For the purpose of listing the other
 changes of variables that we have investigated, it is useful to represent a 
block and its corresponding change of variables by a diagram 
in the following way:
\begin{itemize}
\item The variables involved in the transformation are written explicitly. The legs associated 
with the initial variables appear as thick lines. 
The legs associated with the final variables --which correspond to the invariants that 
enter into the expression of
specific propagators--  are shown as dashed lines. 
\item A blob stands for a branch of legs of which total momentum
parametrizes the change of variables related to the block. 
\end{itemize} 

\begin{figure}
\center
\subfigure[SB A]{\includegraphics[width=7cm]{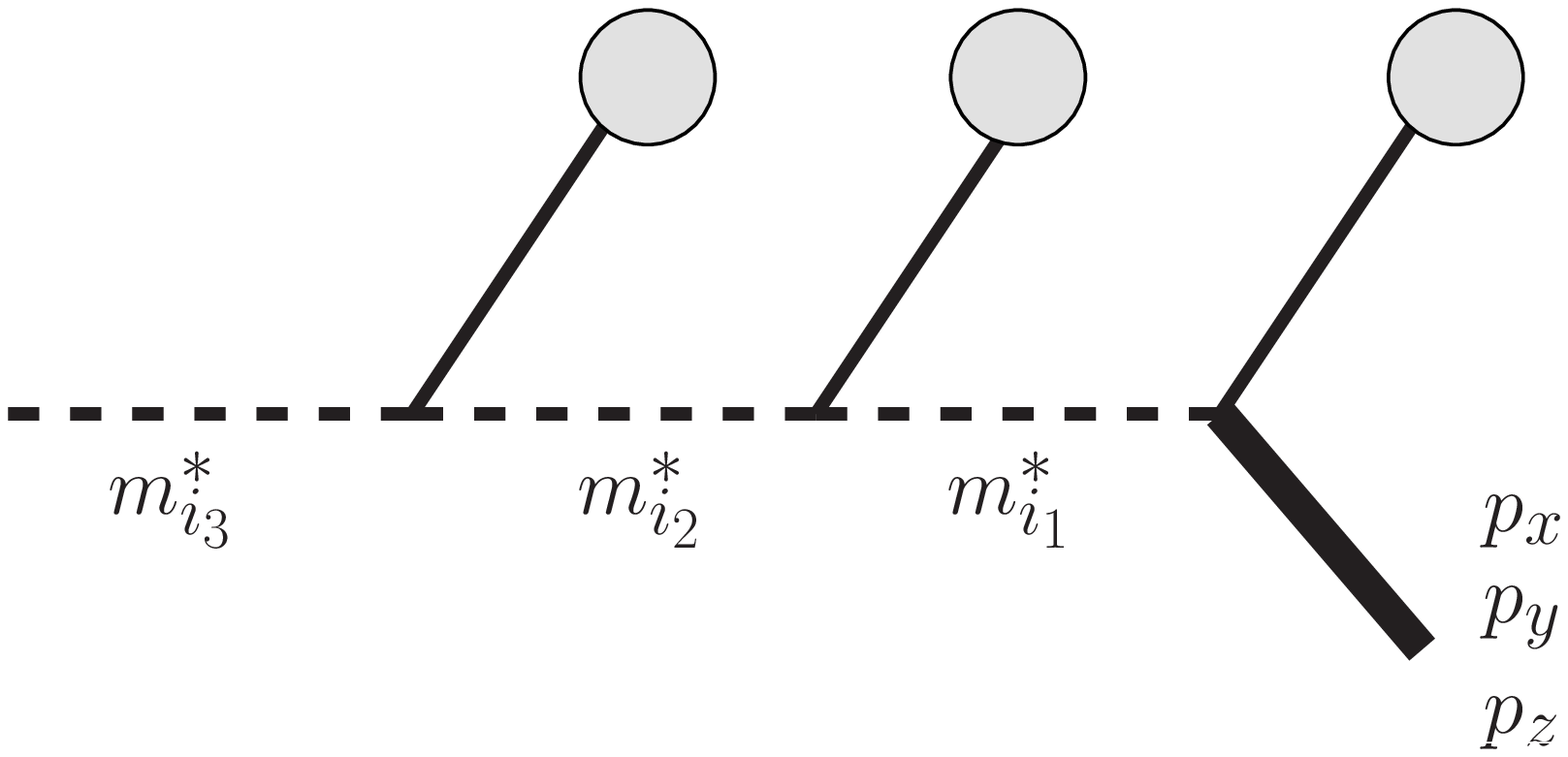}}
\subfigure[SB B]{\includegraphics[width=7cm]{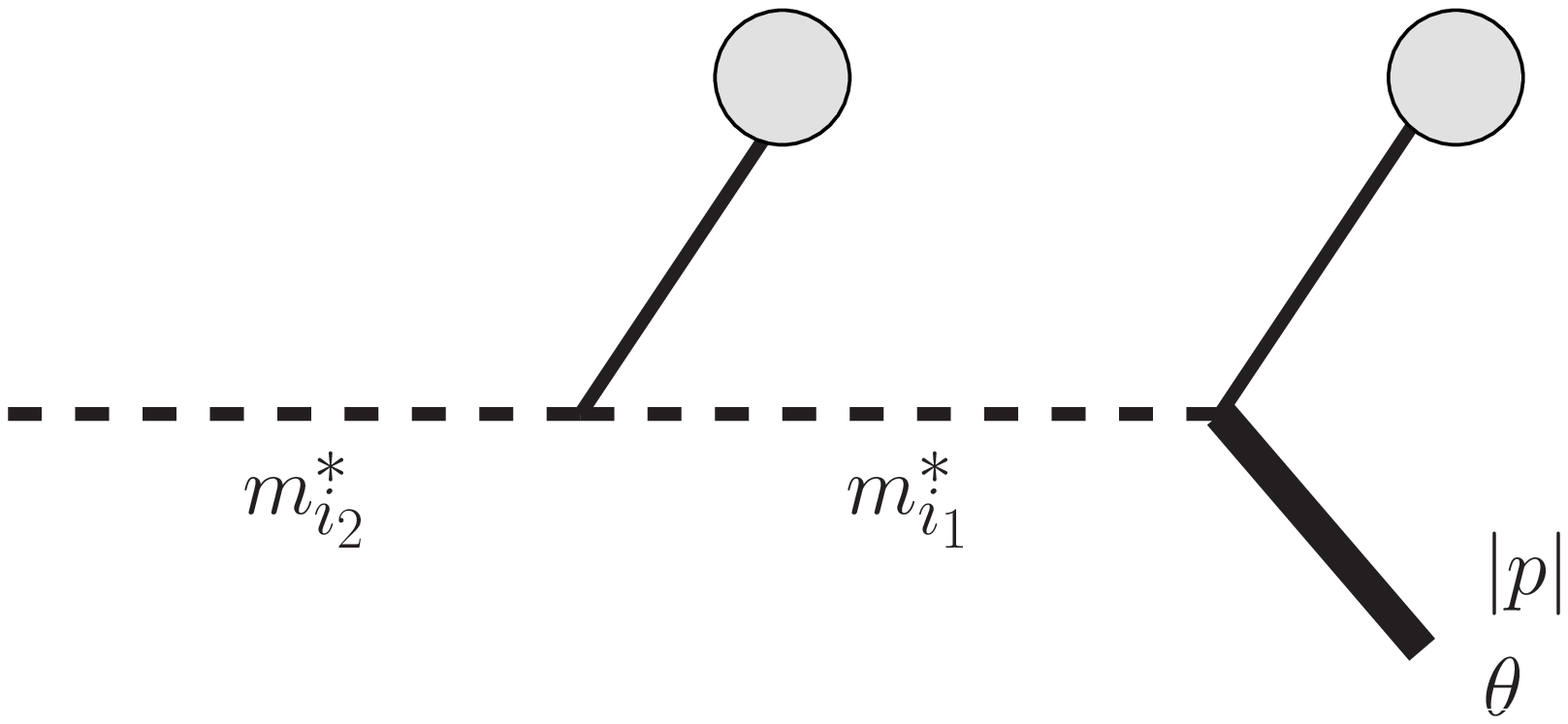}}
\subfigure[SB C/D]{\includegraphics[width=7cm]{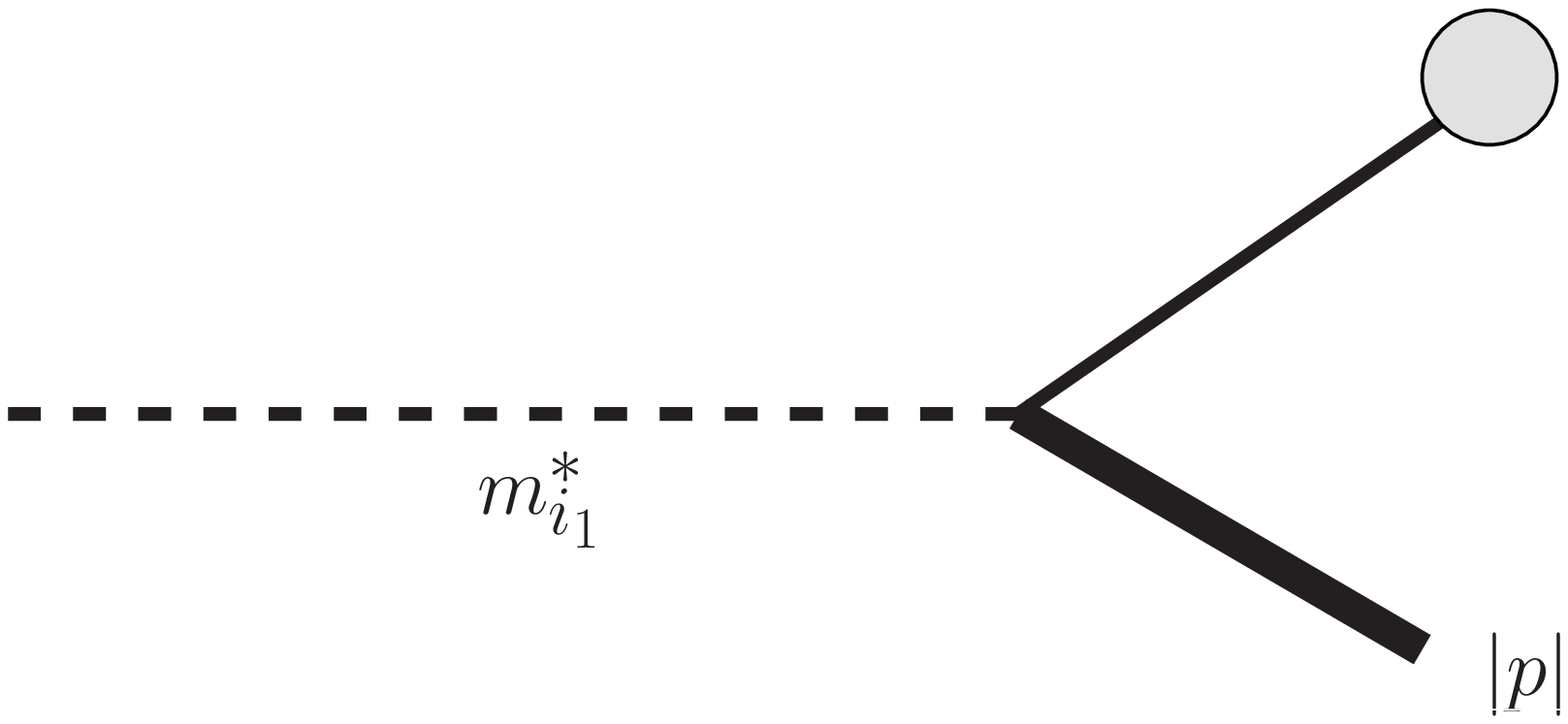}}
\subfigure[SB E]{\includegraphics[width=7cm]{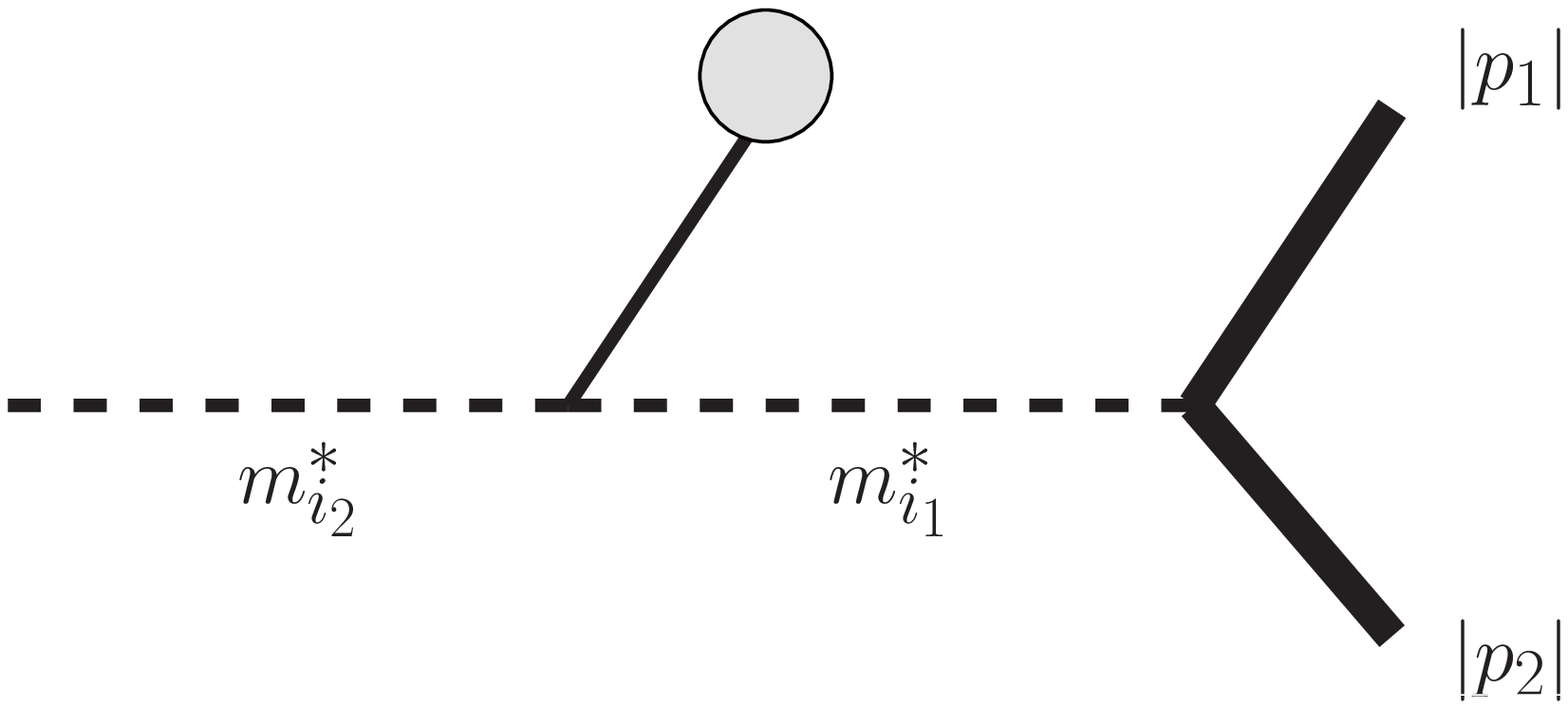}}
\caption{The four secondary blocks with their corresponding 
change of variables.
Initial (resp. final) variables are written explicitly,
and the corresponding legs are represented by thick lines (resp. dashed lines).
 \label{fig:secondary_blocks}}
\end{figure}
In this representation, a blob can a priori be itself decomposed into several secondary blocks.
However, as in the case of the main block,  the change of variables associated with a given secondary block
is only parametrized by the total momentum of each branch represented by a blob,
it does not depend on the details of these branches.
The number of implemented blocks in our algorithm is reduced by our requirement of analytically invertible changes of variables. 
These blocks are displayed in Figure~\ref{fig:secondary_blocks}.
The changes of variables associated with each block are discussed in 
Appendix~\ref{appendix:blocks}. 

\begin{description}
\item[SB A.]  The transformation removes  the 3-momentum of a missing 
particle from the set of integration variables in the parametrization of the phase-space measure.
The new integration variables are the  Lorentz invariants
 $m_{i_1}^*$, $m_{i_2}^*$ and $m_{i_3}^*$  associated with the first, second and third 
 mother particles of this missing particle.

\item[SB B.]  The transformation removes  the energy and the polar angle of a missing 
particle from the set of integration variables in the parametrization of the phase-space measure.
The new integration variables are the  Lorentz invariants
 $m_{i_1}^*$ and $m_{i_2}^*$ associated with the first and second  mother particles of this missing particle.

\item[SB C/D.]  The transformation removes  the energy of a missing 
particle from the set of integration variables in the parametrization of the phase-space measure
(version C).
The new integration variable is the  Lorentz invariant
 $m_{i_1}^*$ associated with the  mother particle of this missing particle.
In version D of this block, the missing particle is replaced by a visible particle, but the
transformation remains the same one.

\item[SB E.] The transformation removes  the momenta $|\bs p_1|$ 
and $|\bs p_2|$ of two visible particles produced by the same resonance
from the set of integration variables in the parametrization of the phase-space measure.
The new integration variables are the  Lorentz invariants
 $m_{i_1}^*$ and $m_{i_2}^*$ associated with the first and second mother particles of these 
 visible particles. The corresponding change of variables is invertible analytically only if at least one of the 
two visible particles is massless.
\end{description}

\begin{figure}
\center
\includegraphics[scale=0.4]{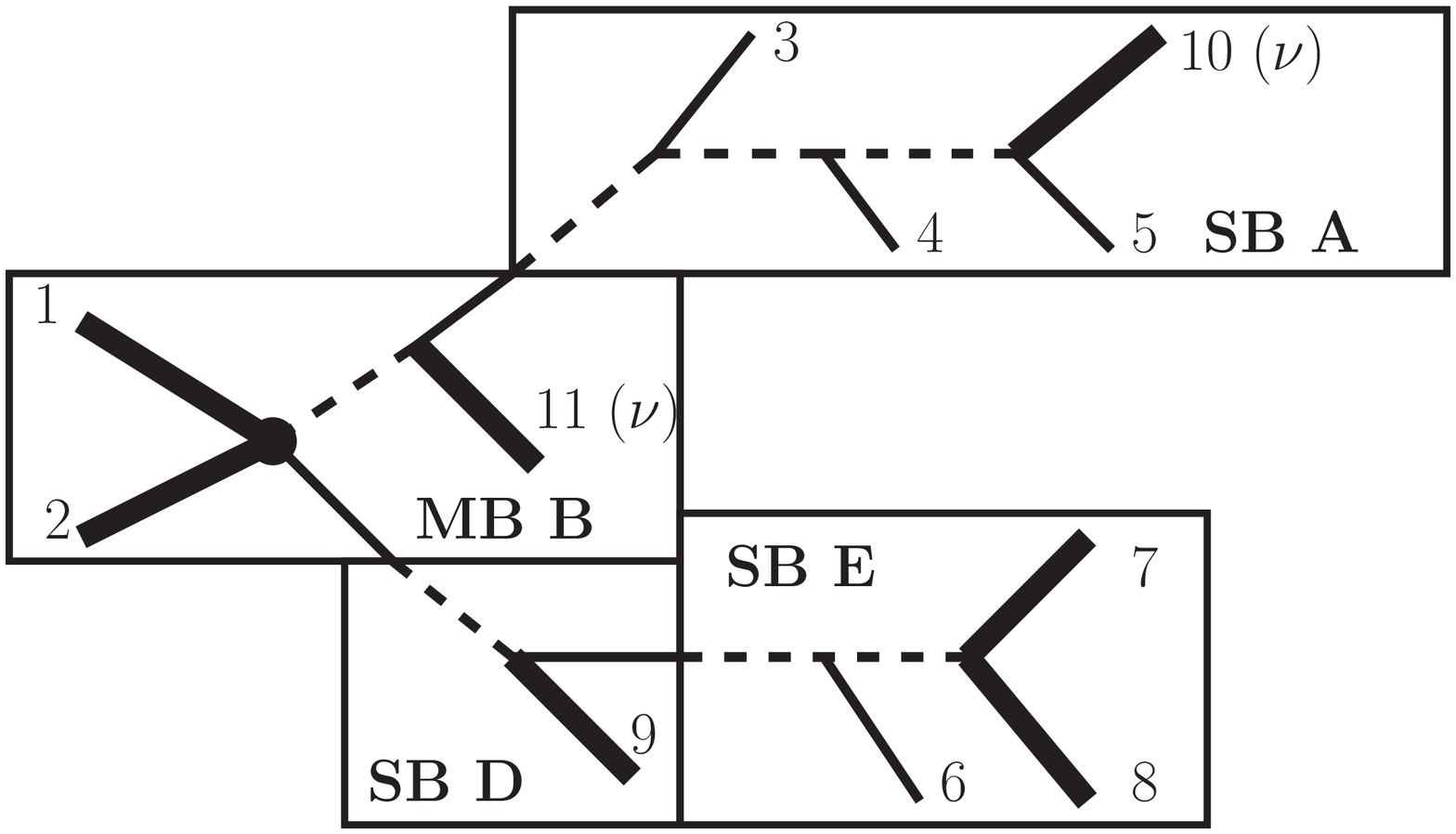}
\caption{Illustration of the structure in blocks optimizing the parametrization of the phase-space 
measure in the case of a specific decay chain.
The missing particles are indicated by the Greek letter $\nu$.
\label{fig:block_composition_ex}
}
\end{figure}

This completes the description of the blocks that can be used in our procedure to optimize the parametrization
of the phase-space measure for the computation of the weights. As a example, we illustrate a composition in blocks in Figure~\ref{fig:block_composition_ex} in the case of a specific decay chain.

\subsection{One-channel phase-space generator}
\label{sec:generator}

Given an optimized phase-space mapping defined by its structure in  
blocks, one can then consider a phase-space generator built upon
this phase-space mapping.
The generation of an arbitrary phase-space point proceeds in two steps:
1)  the generation of the integration variables appearing in the optimized 
parametrization of the phase-space measure, 2) the determination of the momentum
of each leg in the decay chain and the computation of the Jacobian factors.

Concerning the first step, any variable of integration associated with the new phase-space 
mappings introduced in the previous section
enters into one of the three following categories:
\begin{enumerate}

\item The variable controls the strength of a resolution function. 
If the resolution function is a $\delta$ distribution, the variable 
is fixed to the value associated with the experimental event.
Otherwise, the grid of VEGAS is adapted such that the variable is generated according 
to a probability density that reproduces approximately the shape of the resolution function. 

\item The variable controls the strength of a propagator enhancement.
In this case, the variable can be generated according to a probability density
that reproduces exactly the shape of the propagator by using the inverse primitive
function of a Breit-Wigner. 

\item The variable is either the polar or the azimuthal angle of a missing particle.
In this case, the variable is generated according to a uniform distribution
in the interval $[0,\pi]$ or $[0,2 \pi]$ at the first iteration.  
The grid is adapted at each iteration to approximate the optimal probability density.
\end{enumerate}

Once the integration variables have been generated, the kinematics 
of the whole decay chain and the Jacobian factors are computed. 
For each block, the formulas that give the expression of the external momenta
as a function of the variables of integration are discussed in the Appendix.
These formulas are parametrized by the momentum of the branches represented 
by the blobs that appear in the graphical representation in Figures~\ref{fig:main_blocks} 
and \ref{fig:secondary_blocks}. For this reason, one needs to fill the kinematic 
variables in each block in a specific order, starting with the secondary blocks
at the very end of the decay chain, and ending with the main block.

This procedure is best illustrated with the example in Figure~\ref{fig:block_composition_ex}. 
A phase-space point is defined by generating all the integration variables
in the transformed expression of the phase-space measure: the invariant mass of each leg shown as
a dashed line, the direction $(\theta, \phi)$ of any visible particle, 
and the energy of the visible particles represented by the solid thin lines.
Then all other kinematic variables are determined as a function 
of the generated variables, first in the secondary blocks
A and E, then in the secondary block D (by means of 
formulas that are parametrized by the kinematics in block E),
and finally in the main block B (by means of formulas that are 
parametrized by the kinematics of all the secondary blocks).
Such an approach can be easily generalized to the case of an arbitrary decay chain.

One difficulty in our approach is that the boundary of the physical 
phase-space volume cannot be translated into 
simple conditions on the variables of integration. In order to implement
the boundary, we simply check point-by-point in the phase-space that 
 the variables in the main block are  physical (for example, the Bjorken fractions 
$q_1$ or $q_2$ cannot be larger than one), otherwise we throw away the phase-space point.
In some cases, the fraction of unphysical points that are removed in this way may be large. 
Still, the algorithm is rather fast since the generation of a phase-space point is in general much less time-consuming than the evaluation of the squared matrix element.  

\subsection{Multi-channel phase-space generator}
\label{sec:multi_channel}

If a given parametrization of the phase-space measure maps all the peaks in the integrand
simultaneously, an adaptive Monte Carlo integration using only this channel is expected
to be efficient.
But most of the time, each peak in the integrand cannot be mapped onto
a variable of integration in a single phase-space mapping, since the number of peaks is larger 
than the dimension of the phase space.\footnote{This situation could also occur if one of the required 
blocks to build such a 
phase-space mapping corresponds to a change of variables that 
cannot be inverted analytically and hence has not been considered
in our algorithm.} In these cases, we keep 
several channels, \textit{i.e.}
several phase-space parametrizations $\bs{z} \rightarrow \bs{z}'=\bs{P}_i(\bs{z})$
such that each peak in the integrand is mapped onto 
a variable of integration in at least one channel.  
The total integration can be carried out 
using a multi-channel integration approach 
in which every channel $i$ comes with
a phase-space-dependent weight $\beta_i(\bs{z})>0$ in the global parametrization of the phase-space measure:
\begin{equation}
\bs{z} \rightarrow \bs{z}' =\bs{P}(\bs{z})=\sum_i \beta_i(\bs{z}) \bs{P}_i(\bs{z}) ,
\label{eq:multi-channel}
\end{equation}
with the condition $\sum_i \beta_i(\bs{z})=1$.
Each weight $\beta_i(\bs{z})$ must be chosen such that it is significant 
in the phase-space region where the corresponding channel $\bs{P}_i(\bs{z})$
is relevant. In the case of the computation of total cross sections,
this condition can be automatically fulfilled by setting $\beta_i(\bs{z})$ to be proportional
to the amplitude squared of a single diagram associated with the  channel $i$~\cite{Maltoni:2002qb}.
In analogy to the single-diagram enhanced method, we choose to set the weight $\beta_i(\bs{z})$
to be proportional to the product of the peaks that are mapped onto
integration variables in the corresponding phase-space mapping.

In comparison with previous implementations 
for the evaluation of the matrix element weights,
this multi-channel approach is expected to speed up the convergence of the integration,
especially in the case of an over-constrained topology.
An example of such a topology has been investigated in~\cite{Abazov:2006bd,Abulencia:2007br,Abazov:2004ym},
where either the helicity of the W boson or the mass of the top quark 
is reconstructed from $t \bar t$ events in the semi-leptonic
channel. In these analyses, a single channel was used for the evaluation of the weights,
leaving unmapped a subset of peaks in the integrand.
On the contrary, our procedure always maps a given peak in the integrand onto a variable of integration
in at least one channel.

The whole procedure that we have presented so far has been implemented in the MadGraph
framework, and the corresponding module has been named MadWeight.
For a given decay chain and a transfer function for the final state objects,
the optimized phase-space mappings are automatically selected, and the resulting
multi-channel phase-space generator is used for the evaluation of the weights.
While this procedure applies for virtually all cases,
the speed of convergence of the numerical integration strongly 
depends on the process under investigation, and whether the calculation
time is a serious limitation or not has
 to be assessed on a case-by-case basis.

\subsection{Validation of the phase-space generator}

\label{sec:checks}

\begin{table}[t]
\center
\begin{tabular}{|c|c|c|c|}
\hline
 $l$ & blocks & integrated volume  \\
\hline \hline
3 & MB A             &  $6.30\times 10^{-5}$  \\
3 & MB B             &  $6.30\times 10^{-5}$ \\
3 & MB C             &  $6.30\times 10^{-5}$ \\
6 & MB D             & $694$  GeV$^6$        \\
4 & MB E             & $0.0166$ GeV$^2$    \\
4 & MB F             & $0.0166$ GeV$^2$    \\
5 & MB B + SB A   & $3.89$ GeV$^4$     \\
4 & MB B + SB B   & $0.0166$ GeV$^2$ \\
3  & MB B + SB C  &  $6.30\times 10^{-5}$\\
3  & MB B + SB D  &  $6.30\times 10^{-5}$\\
4 & MB B + SB E   &  $0.0166$ GeV$^2$  \\
\hline
\end{tabular}
\caption{Phase-space volumes $\int dq_1 dq_2 d\phi_n 1/(s q_1q_2)$ for $l$ massless particles produced
 in hadron-hadron collisions at $\sqrt{s}=1$ TeV.  The number $l$ of final-state particles 
is indicated in the first column.
The second and third columns indicate the structure in blocks
defining the phase-space mapping that is used
to calculate the volume with our phase-space generator, and the numerical value that we obtained.
Each number is in agreement with the exact value of the phase-space volume at three digit accuracy. 
}
\label{tab:PSvolume_massless}
\end{table}

One potential issue related to our phase-space mappings optimized 
for the computation of the weights is the fact that some of the 
associated Jacobians develop  singularities in specific 
phase-space regions. These singular regions are an artefact 
of the change of variables. In our case they have a null measure
in the integration volume. One can therefore split the integration volume
into a volume $V_1$ where the Jacobian is finite
and a volume $V_2$ that contains the singular region and that can be made
arbitrary small compared to the volume $V_1$. At any given accuracy, 
we can ignore the contribution from the volume $V_2$ provided
that $\epsilon=V_2/V_1$ is sufficiently small. At the numerical level though,
one may fear that  instabilities will appear in this procedure. 

\begin{table}[t]
\center
\begin{tabular}{|c|c|c|c|}
\hline
 $l$ & blocks & integrated volume \\
\hline \hline
3 & MB A             & $3.49 \times 10^{-5}$  \\
3 & MB B             & $3.49 \times 10^{-5}$  \\
3$^*$ & MB C             & $4.13 \times 10^{-5}$ \\ 
6 & MB D             & $124$ GeV$^6$        \\
4 & MB E             & $8.17\times 10^{-3}$ GeV$^2$ \\ 
4 & MB F             & $8.17\times 10^{-3}$ GeV$^2$  \\
5 & MB B + SB A   & $1.28$  GeV$^4$           \\
4 & MB B + SB B   & $8.17\times 10^{-3}$ GeV$^2$ \\  
3  & MB B + SB C  & $3.49 \times 10^{-5}$  \\
3  & MB B + SB D  &  $3.49 \times 10^{-5}$ \\
4$^*$ & MB B + SB E   & $9.78$  GeV$^2$ \\

\hline
\end{tabular}
\caption{Phase-space volumes $\int dq_1 dq_2 d\phi_n 1/(s q_1q_2)$ for $l$ particles
with a mass $m=50$ GeV produced
 in hadron-hadron collisions at $\sqrt{s}=1$ TeV.  The number $l$ of final-state particles 
is indicated in the first column. A star $^*$ indicates that the mass of one of the final state 
particles is set to zero, as this condition is required by one of the blocks. 
The second and third columns indicate 
the structure in blocks
defining the phase-space mapping
 that is used
to calculate the volume with our phase-space generator, and the numerical value that we obtained.
Each number is in agreement with the exact value of the phase-space volume at three digit accuracy. 
}
\label{tab:PSvolume_massive}
\end{table}

In practice, we have not encountered any numerical instabilities resulting from a change of variables
that is associated with a specific phase-space block. Any phase-space block and the related change of variables that have been defined in our procedure have been checked by reproducing the volume of the entire phase-space
region with our phase-space generator using a parametrization of the phase-space measure that involves this block.
This Monte Carlo procedure to compute the phase-space volume has a very poor convergence, as the
phase-space mappings that are optimized for the computation of the weights are clearly inefficient for the
computation of just the phase-space volume. Nevertheless, by increasing the number of generated phase-space points,
we checked that the phase-space volume is reproduced with an accuracy better than one percent for
each tested phase-space mapping. We first set the mass of the final-state particles to zero and obtained the results summarized in Table~\ref{tab:PSvolume_massless}. We then considered the case of massive particles in the final state and obtained the results summarized in Table~\ref{tab:PSvolume_massive}.

In order to validate the multichannel implementation, we also computed the total cross section 
of several processes by integrating the squared matrix element with our phase-space generator.
This can be achieved by setting all transfer functions to one. Here again, the convergence of the 
numerical integration is poor, as the phase-space parametrization is not designed for such computation.
By using a very high statistics, we reproduced the total cross sections associated with the processes
listed in the first column of Table~\ref{checksXsections}.

\begin{table}[t]
\center
\begin{tabular}{|c|c|c|c|}
\hline
 process & $\sigma^{\textrm{MW}}/\sigma^{\textrm{ME}}$ &  channels & blocks \\
\hline \hline
$pp \rightarrow (W \rightarrow jj)j$ & 0.982(6) & 3 & MB A \\
$pp \rightarrow (W \rightarrow l\nu)$ & 0.9991(14) & 1 & MB B \\
$pp \rightarrow [W\rightarrow  \tilde{\nu}_\tau (\tilde{\tau}^->\tau^- \tilde{\chi})]$ &
1.003(5) & 1 & MB B; SB C \\
$pp \rightarrow 2[\tilde{\mu} \rightarrow \mu \tilde{\chi}]$ & 1.020(5) & 3 & MB B,F; SB C  \\
$pp \rightarrow 2[t \rightarrow b (W \rightarrow l \nu_l)]$ & 1.000(25) & 1 & MB D \\
$\begin{array}{c} pp \rightarrow [t \rightarrow b (W^+ \rightarrow l \nu_l)] 
+[\bar t \rightarrow \bar b (W^- \rightarrow jj)]
\end{array}$  &  0.94(5) & 6 & MB B, SB D,E \\
$pp \rightarrow h \rightarrow (W^+ \rightarrow \mu^+ \nu_m)(W^- \rightarrow \mu^- \bar \nu_m )$ & 0.99(2) & 1 & MB E \\
\hline
\end{tabular}
\caption{Validation of the phase-space generator by computing total cross sections.
The processes under consideration are written in the first column. 
The second column gives the ratio of the cross section computed with MadWeight over the one computed with 
MadEvent~\cite{Maltoni:2002qb}. The third column indicates the number of channels that are used in the 
MadWeight integration, and the last column indicates the blocks that are involved in that integration.
}
\label{checksXsections}
\end{table}

\section{Example of applications}
\label{applications}

In this section we illustrate a few examples of studies that can be achieved with MadWeight. The following analyses are based 
on simulated events generated with MadGraph/MadEvent~\cite{Alwall:2007st}. The events are passed through Pythia~\cite{Sjostrand:2006za}
for the showering and the hadronization. Electrons and muons are assumed to be reconstructed with 100\% efficiency and with excellent resolution if they have a pseudo-rapidity $|\eta|<2.4$. Detector response simulation is performed using PGS~\cite{PGS}
 which takes into account geometrical acceptance, finite granularity and energy resolutions of typical calorimeters used in LHC experiments. Jets are then reconstructed 
based on the $k_t$ algorithm~\cite{Catani:1993hr, Ellis:1993tq, Cacciari:2005hq}
and applied on the calorimeter cells fired by the generated stable or quasi-stable particles. 

The  transfer functions  $W_i^E(x^i,y^i)$, Eq.~(\ref{eq:TF2}) with $i$ running over all reconstructed jets are determined from an independent $t\bar{t}$ sample where well separated jets (including light and b jets) are matched to the corresponding partons.
We consider a double-Gaussian shape function characterized by 5 parameters: the means and the widths
of the two Gaussian distributions, and their relative normalization.
We fit these five parameters in each 20 GeV bin in jet energy from 40 GeV to 200 GeV.
The energy dependence of the mean and the width of each Gaussian distribution is then
approximated by the parametrization $c_1+c_2\sqrt{E}+ c_3 E$, with the coefficients $c_1$, $c_2$
and $c_3$ extracted 
from a $\chi^2$ fit to the values of the four parameters of Gaussian distributions 
in each energy bin. The relative normalization of the two Gaussian distributions
is assumed to be energy independent, and is fixed to the average of the
corresponding values in each energy bin.
The typical resolution for jet energy is between 5 and 12 GeV, with tails parametrized by Gaussian of variances as large as 30 GeV.  

\subsection{Top-quark mass measurement}
\label{sec:top_mass}

The top-quark mass measurement by means of the matrix element method was published for the first time by the 
D$\emptyset$ collaboration using the single-leptonic final state arising from top-quark pair production~\cite{Abazov:2004cs}. The method has been later extended also to include a simultaneous determination of the Jet-Energy-Scale uncertainty. 
The  accuracy of the experimental determination of $m_t$ has been further improved by the
contribution of other studies based on matrix element method~\cite{Abulencia:2007br,Abulencia:2006ry,Abazov:2006bg}. In these analyses, a dedicated phase-space integration was performed to define the event weight. Our algorithm provides this weight automatically based on the blocks given in Table~\ref{checksXsections}, last but one line.

\begin{figure}
\subfigure[]{\includegraphics[scale=0.61]{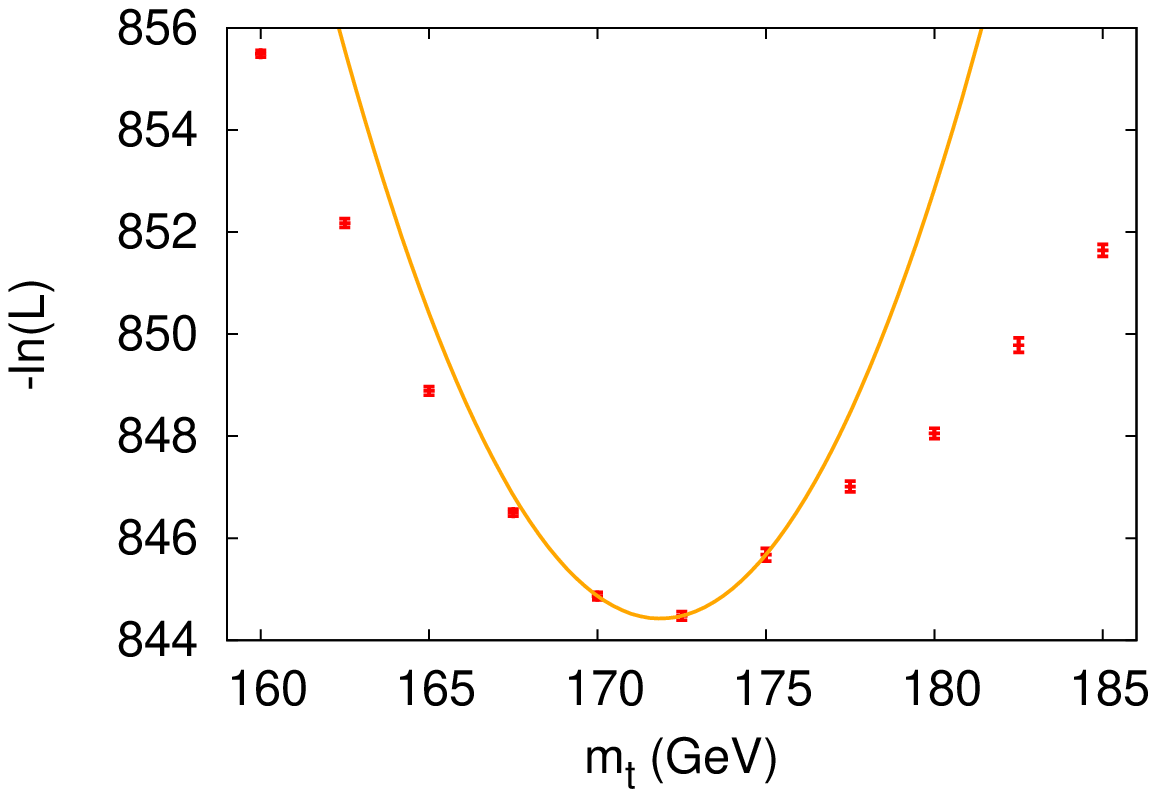}}
\subfigure[]{\includegraphics[scale=0.61]{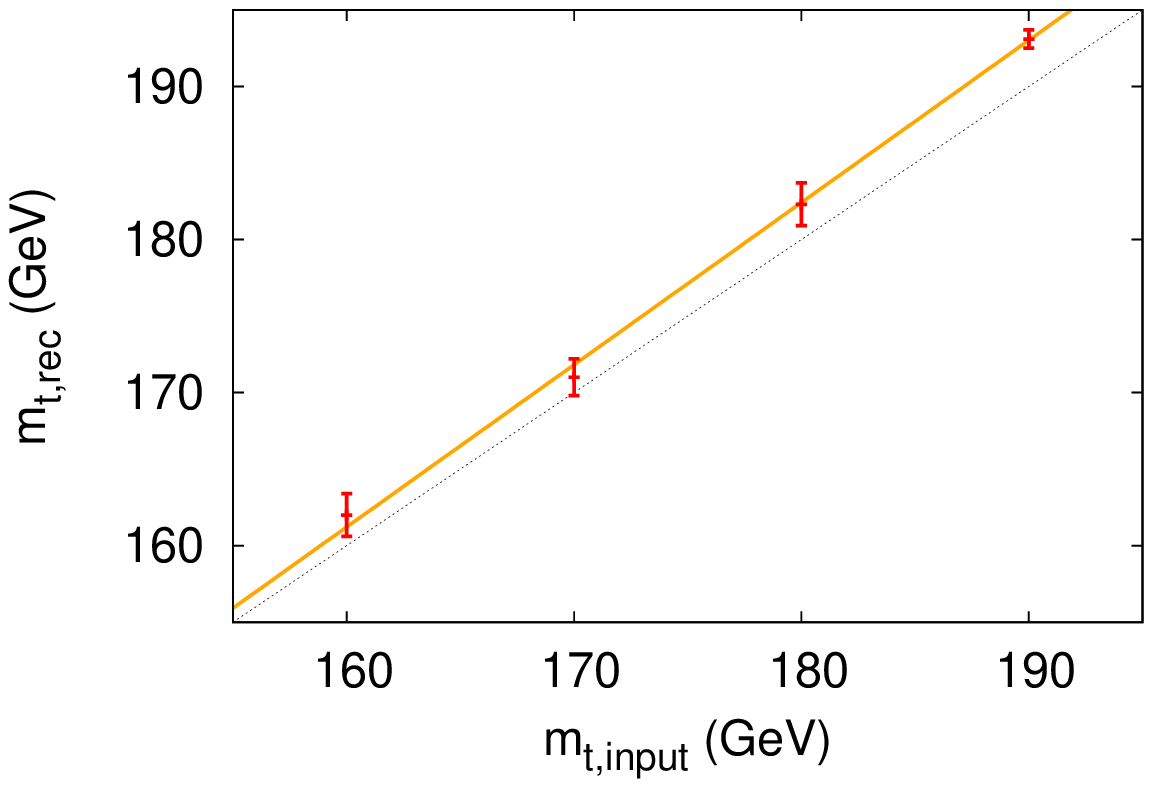}}
\caption{
 (a) Logarithmic likelihood values for a sample of $20$ events generated with $m_{t, \textrm{ input}}=170$ GeV.
The solid line is a parabolic fit to the points near the minimum.
The statistic error is estimated by
the half width of the distribution at $\log (L/L_{\textrm{max}})=0.5$ and is extracted from the fit.
(b) Calibration of the matrix element mass fitting procedure. The errorbars correspond to the
 the value of mass of the top quark and the associated statistic error reconstructed from $t \bar{t}$ samples generated with different input values of $m_t$.
The solid line is a linear fit to the four points and the dotted line corresponds to $m_{\textrm{rec}}=m_{\textrm{input}}$.}
\label{plot_topmass}
\end{figure}

As an example of application of our automatic reweighting algorithm, we illustrate the performance of the method for the determination of the top-quark mass at the LHC, by using a small statistics of $t \bar{t}$ events in the single lepton final state: 
\begin{equation}
pp \rightarrow [ \bar t \rightarrow \bar b (W^- \rightarrow \mu^- \bar \nu_{\mu}] [t \rightarrow b (W^+ \rightarrow jj)].
\end{equation} 
For the sake of simplicity, we assume that there is no background and $20$ signal events after selection. Pseudo-data have been simulated with an input top-quark mass at $170$ GeV.
The selection requires one muon  with a reconstructed transverse momentum above $10$ GeV and exactly four isolated jets with a reconstructed transverse momentum above $20$ GeV.

The determination of the top-quark mass from our sample of pseudo-data is obtained by the minimization of $-\log(L)$ with respect of $m_t$ where the likelihood $L$ is defined -up to a normalization factor- by the product of the weights calculated for each event
\begin{equation}
-\log (L)=-\sum_{i=1}^{N} \log[P(x_i;m_t)]\,.
\label{likelihood_log}
\end{equation}
The acceptance of the detector and the cuts imposed on the sample may depend on the input mass of the top quark. Such a dependence might introduce a  bias in the extraction of $m_{\textrm{top}}$ from the fit of the likelihood given in Eq.~(\ref{likelihood_log}). We explicitly tested that  in our 
pseudo-data  this bias is very small and we therefore ignored it in this  example.

The values of $-\log [L(m_t)]$ for different assumptions of $m_t$ are displayed in 
Figure~\ref{plot_topmass}(a). A clear minimum is observed close to the input mass value. A parabolic fit gives the value $m_t=171.9\pm 2.0_{stat}$ GeV. Using ten independent samples generated under 
the same conditions, we extracted the expected $m_t$ value and the expected statistic error for  
such a measurement:  $m_t = 173.5 \pm 3.7 _{stat}$ GeV. We also identified a 
small significant shift between the input and the reconstructed mass by repeating the analysis for other input masses, as can be seen from the calibration curve displayed in Figure~\ref{plot_topmass}(b).  This curve was obtained by generating $t \bar t$ samples with top quark masses of $160$, $170$, $180$ and $190$ GeV with the same selection procedure and the same fitting procedure, and with a statistics of 100 events per sample. 
The small bias between the input and the reconstructed mass of the top quark 
can possibly result from the effect of initial state radiation, which is not
completely removed in our event selection criteria~\cite{Alwall:2010cq}.

\subsection{Spin identification in decay chains with missing energy}

As a second illustration we address  the challenge of determining  the spin of new particles. A simple example is the production of a light charged Higgs boson with a mass close to the $W$ boson mass. The information on the spin of a resonance is passed through the angular distribution of its decay products. If the momentum of each final-state particle produced in the decay chain is measured, the angular distributions can be reconstructed and the spin of the resonance identified. In fact in many cases, such as the one we have chosen, the final state is characterized by missing transverse energy from undetected particles and the angular distributions of the decay products cannot be fully determined. The interesting question becomes therefore whether the available information from the final state is sufficient to discriminate between different spin assignments in the decay chain. The matrix element method appears to be particularly relevant in this case, since the event weight will encompass the whole available event kinematics including the spin correlation effects that survive after the experimental reconstruction of the events. 

In the following example, we assume that the production of the signal and its irreducible background proceeds exclusively via the production of top quark pair that subsequently decay into $H^+ +b$ or into $W+b$, with $ m_{H^{\pm}}\simeq m_{W^{\pm}}$, followed by a leptonic decay of both bosons.
The signal process is
\begin{equation}
pp \rightarrow [t \rightarrow b (H^+ \rightarrow \tau^+ \nu_{\tau})] [ \bar t \rightarrow \bar b (W^- \rightarrow \mu^- \bar \nu_{\mu})],
\label{eq_H+_signal}
\end{equation}
and the corresponding irreducible background results from the production of a pair of $W$ bosons
\begin{equation}\label{eq_H+_bg}
pp \rightarrow [t \rightarrow b (W^+ \rightarrow \tau^+ \nu_{\tau})] [ \bar t \rightarrow \bar b (W^- \rightarrow \mu^- \bar \nu_{\mu})].
\end{equation}
At the reconstruction level, we required the presence of exactly two jets with a $p_T$ larger than $20$ GeV, one $\tau^+$ ---assumed to be reconstructed as precisely as the other charged leptons---
 and one $\mu^-$. The cuts on these leptons are $|\eta|<2.4$ and $p_T>5$ GeV.
We reject the events containing photons or electrons with $p_T>5$ GeV and $|\eta|<2.4$, while no restriction is imposed on the number of jets with a $p_T$ less then $20$ GeV. The transfer functions associated with the other particles have the same parametrization as described in  Section~\ref{sec:top_mass}. We arbitrarily choose a final relative normalization of signal and background events, working with a sample of 240 signal events and 760 background events.

Before proceeding further, we stress that while providing an interesting case study, our example cannot be regarded very realistic. 
First, the relative cross sections and reconstruction efficiencies for the signal and the background have been  chosen arbitrarily. 
Second, this such a light charged Higgs is not favoured by the present constraints, which point to much higher masses. 
Finally, the tau lepton reconstruction is idealized as it is considered here on the same footing as as a muon. A more realistic approach would consist of taking into account the energy loss from the tau decay with a dedicated transfer function for the energy of the tau. Nonetheless, as shown below, this example illustrates quite well the power of the matrix element method.

\begin{figure}
\subfigure[]{\includegraphics[scale=0.58]{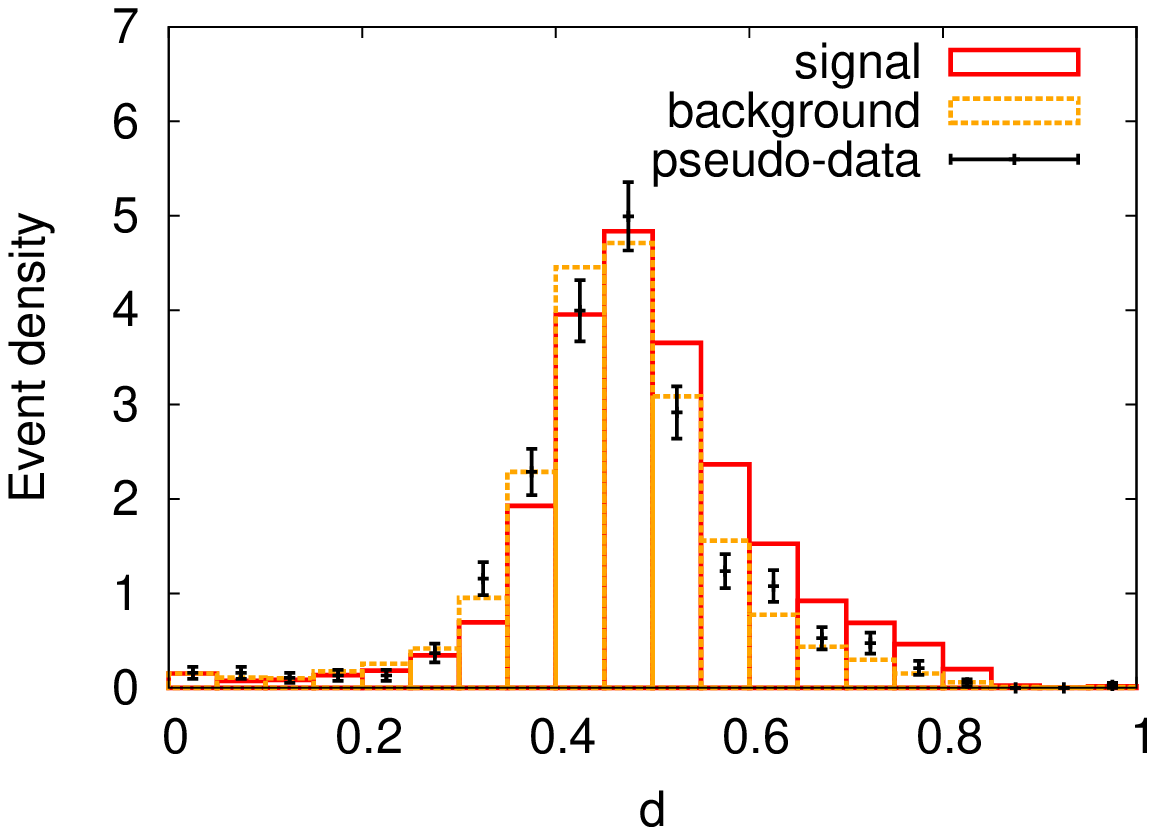}}
\subfigure[]{\includegraphics[scale=0.58]{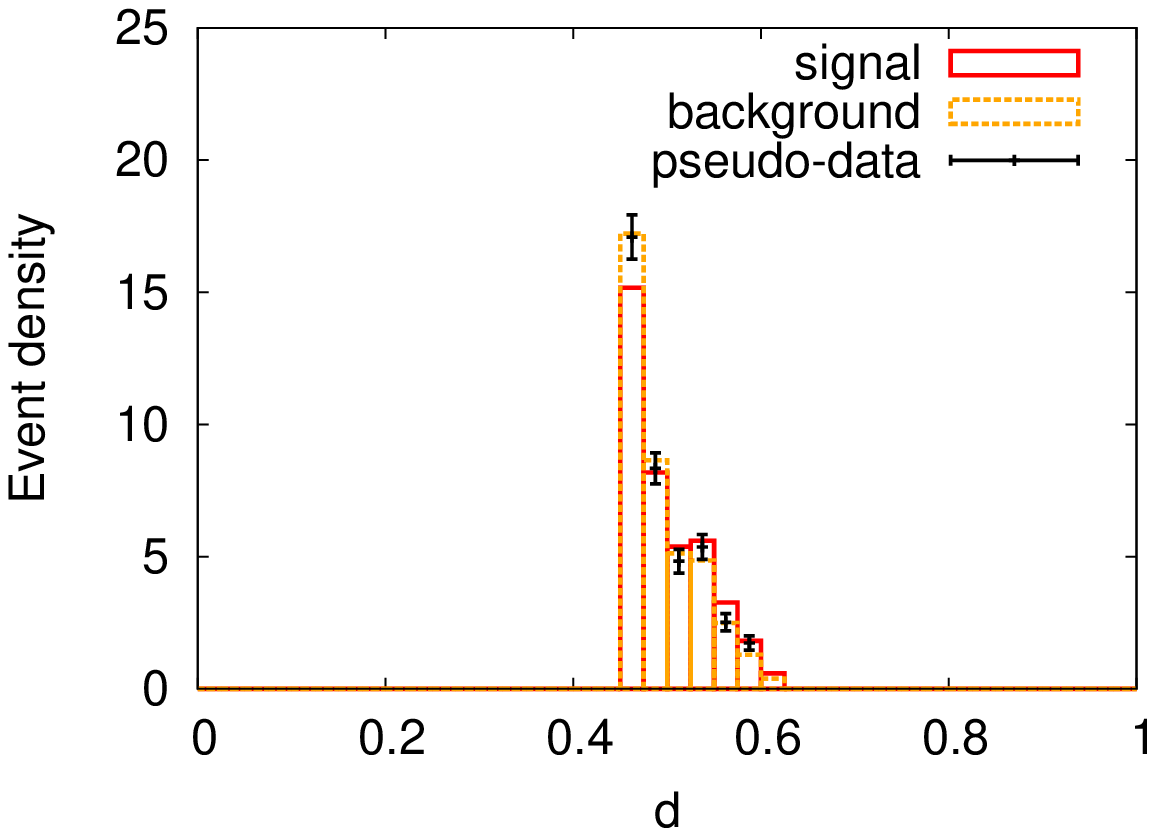}}
\caption{Expected normalized distribution of events with respect to the discriminant $d$ built upon (a) the matrix element weight and (b) the $p_T$ of the tau for a pure signal sample (solid histogram) and for a pure background sample (dashed histogram). The errorbars are the distribution associated with the pseudo-experiment sample, assuming that the statistical error on the number $N$ of events in a bin is given by $\sqrt{N}$.
\label{fig:spin_discriminant}
}
\end{figure}

Let us define $P_S(x)$, $P_{B}(x)$ as the weights evaluated for the event final state $x$ under the signal and the background hypotheses, respectively. These weights can be calculated from the signal and background full matrix element as defined in Eq.~(\ref{weight_def}). Alternatively, they can be associated with a normalized differential cross section with respect to a single observable, such as the $\tau^+$  transverse momentum
\begin{equation}
P_{S,B}(x) \rightarrow \frac{1}{\sigma_{S,B}} \frac{d\sigma_{S,B}}{dp_T} \left[  p_T(\tau^+) \right],
\label{eq:pTweight}
\end{equation}
which also captures the spin effects. The advantage of the weights defined in Eq.~(\ref{eq:pTweight}) is in their simplicity:
their evaluation only requires to use a standard phase-space generator that is optimized for the computation of cross sections. Such an observable, for example,
is very commonly used in the determination of the polarization of the $W$ bosons in top events and provides us with a useful benchmark to study the
increased sensitivity that the matrix element method might provide.

These weights can then be combined to build an event-by-event discriminating variable
\begin{equation}
d(x)=\frac{P_{S}(x)}{P_{S}(x)+ P_{B}(x)}.
\label{eq:definition_discriminant}
\end{equation}
The normalized-to-one distributions of events as a function of the discriminant variable $d$ are shown in Figure~\ref{fig:spin_discriminant} for the two cases. The solid (resp. dashed) histogram is the distribution expected for a pure sample of signal (resp. background) events. These distributions have been generated from large samples, in order to allow us to neglect the statistical fluctuations. Spin correlation effects are expected to give rise to different values for the weights under the two spin hypotheses. Nevertheless, for most of the events, this disparity is expected to be small, resulting in a discriminant close to $d\simeq 0.5$. Note that in our example, signal and background events are characterized by the same topology with intermediate particles of the same mass. Only the spin of the intermediate $W$ or $H$ resonances differ between the two decay chains.  The  distributions clearly shows that the discriminant power is substantially reduced when only the information on the transverse momentum of the $\tau^+$ is retained.

\begin{figure}
\subfigure[]{\includegraphics[scale=0.58]{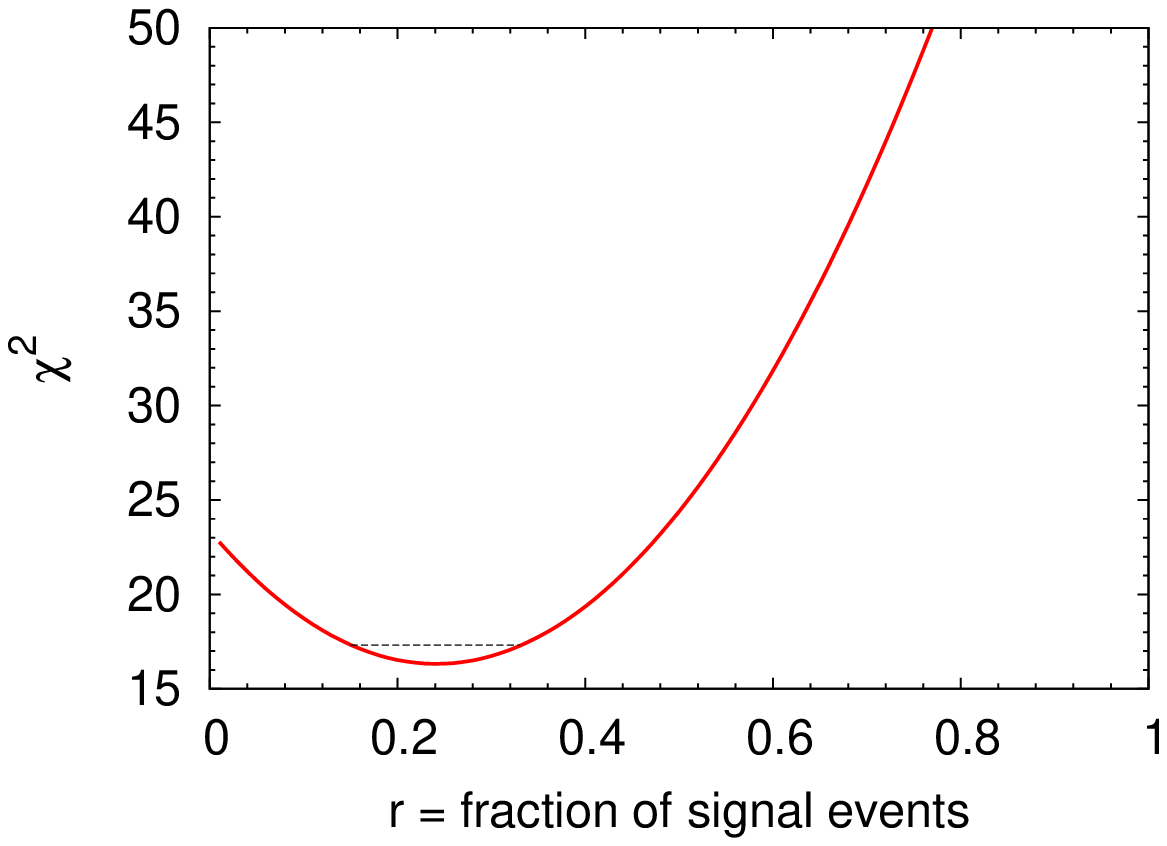} }
\subfigure[]{\includegraphics[scale=0.58]{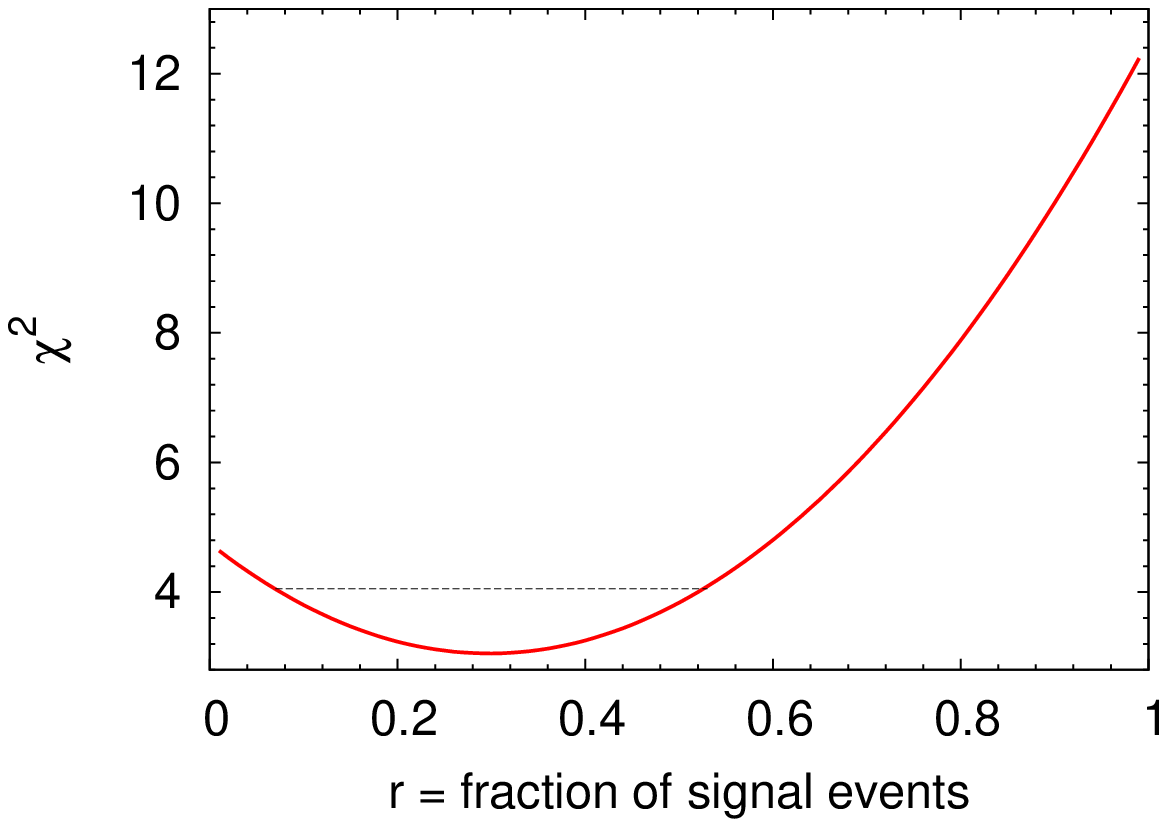}}
\caption{ $\chi^2$ values associated with the fit of the pseudo-experiment data to the theoretical prediction parametrized by the fraction $r$ of signal events for the weights calculated from (a)
the matrix element, (b) the $p_T$ of the tau. Each dashed line represents the one-standard-deviation interval defined by the condition $\chi^2(r)< \chi^2_{\textrm{min}} + 1$.
\label{fig:chi_square} }
\end{figure}

Yet, for some events the discriminant is significantly different of $0.5$, corresponding to configurations clearly  favoured by one of the two hypotheses. Such events influence the shape of the distributions and allow us to distinguish them. One can take advantage of this difference  to find out the fraction of signal events in the pseudo-experiment sample. The normalized-to-one distribution associated with the 
pseudo-experiment sample as a function of  the discriminant variable $d$  is also displayed in  Figure~\ref{fig:spin_discriminant}. The fraction of signal events in the pseudo-experiment sample can be  reconstructed by a 
least-square fit, i.e. by minimizing  
\begin{equation}  
\chi^2(r)=\sum_{\textrm{bins in } d} \frac{ \left\{ \mathcal{P}_{data}(d)- [r \mathcal{P}_S(d)+(1-r)\mathcal{P}_{B}(d) ]\right\}^2}{[ \Delta \mathcal{P}_{data}(d)]^2}
\end{equation}
where  $\mathcal{P}_{S}$, $\mathcal{P}_{B}$,  
 are the expected binned distributions for signal and background events and  $\mathcal{P}_{data}$ is the binned distribution 
associated with the pseudo-experiment sample.

The $\chi^2$ values as a function of the fraction of signal events are shown in Figure~\ref{fig:chi_square}. The best fit for each discriminant are obtained for $r=24\pm9\%$ and $r=30\pm 23 \%$, respectively. Both results are compatible with the true fraction of signal events, but the discriminant using matrix elements increases the accuracy by more than a factor of two.

\subsection{Smuon pair production at the LHC}

Over the past fifteen years, a tremendous amount of work has been devoted to new techniques for mass reconstruction of new particles that might be produced at the LHC. According to most scenarios, the hypothetical new physics states are not expected to be directly observed experimentally, {\it  i.e.}, they appear as intermediate states in specific decay chains or they escape from the detector without interacting with it. Their mass can hence only be reconstructed indirectly, by making a number of assumptions on the decay chain at work. The number of assumptions in turn is directly correlated to the amount of information that can be extracted from the decay chain. However, due to the lack of constraints on physics beyond the standard model, the proposed techniques have to be general enough, at least if they are aimed at reconstructing the mass of new hypothetical particles in the early stages of investigation. Further more, the limited knowledge of the detector has to be taken into account. It is in this context that a number of mass measurement techniques based on kinematic methods have been proposed in the literature. They can be classified according to the type of decay chains that they address and according to the assumptions on which they rely~\cite{Barr:2010zj}.

\begin{figure}
\center
\includegraphics[scale=0.55]{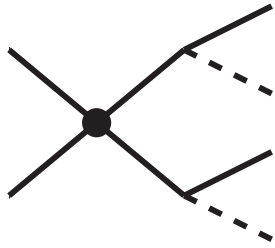}
\caption{Generic decay chain corresponding to the production of two resonances followed by their decay into  weakly-interacting and
 standard model particles. The weakly-interacting particles are represented by the dashed lines.
\label{fig:short_decay_chain}}
\end{figure}

Despite the plurality of kinematic variables that have been proposed, mass determination remains very challenging for specific decay chains. One well-known example of a difficult topology is the production of two  resonances followed by their decay into a weakly-interacting and
a standard model particles, shown in Figure~\ref{fig:short_decay_chain}. In this case, the kinematic methods that have been proposed to reconstruct simultaneously the mass of the two new particles require a very high statistics. Whether their sensitivity is sufficient  under real experimental conditions still remains to be determined. 
A complementary way to address the same  problem is to ask what would be the maximum sensitivity achievable, given a very detailed set of hypothesis to be tested that not only include masses but also the spin and couplings information,  {\it  i.e.}, taking into account the full theoretical model prediction. 

The problem can be investigated with the matrix element method that usually makes use of the strongest assumptions on the analysed events~\cite{Cranmer:2006zs}. One way to dramatically increase the theoretical information is to assume that the masses of the new physics states are the only unknown properties of the decay chain. We therefore consider a specific decay chain corresponding to the topology in Figure~\ref{fig:short_decay_chain}: the production of a pair of smuons followed by their decay into a muon and a neutralino
\begin{equation}
\label{susyDC}
pp \rightarrow ( \tilde{\mu}_r^+ \rightarrow \mu^+ \tilde{\chi}_1) (\tilde{\mu}_r^- \rightarrow \mu^- \tilde{\chi}_1 )\, .
\end{equation}
We suppose that we have isolated a pure sample of events that correspond to the decay chain in Eq.~(\ref{susyDC}). Further, we assume a perfect reconstruction of the kinematics of the two muons in each event. Within these assumptions, the significance that can be achieved with the matrix element method provides us with an upper bound on the significance that can be delivered by any realistic analyses at a given luminosity. 

\begin{figure}
\subfigure[]{\includegraphics[scale=0.58]{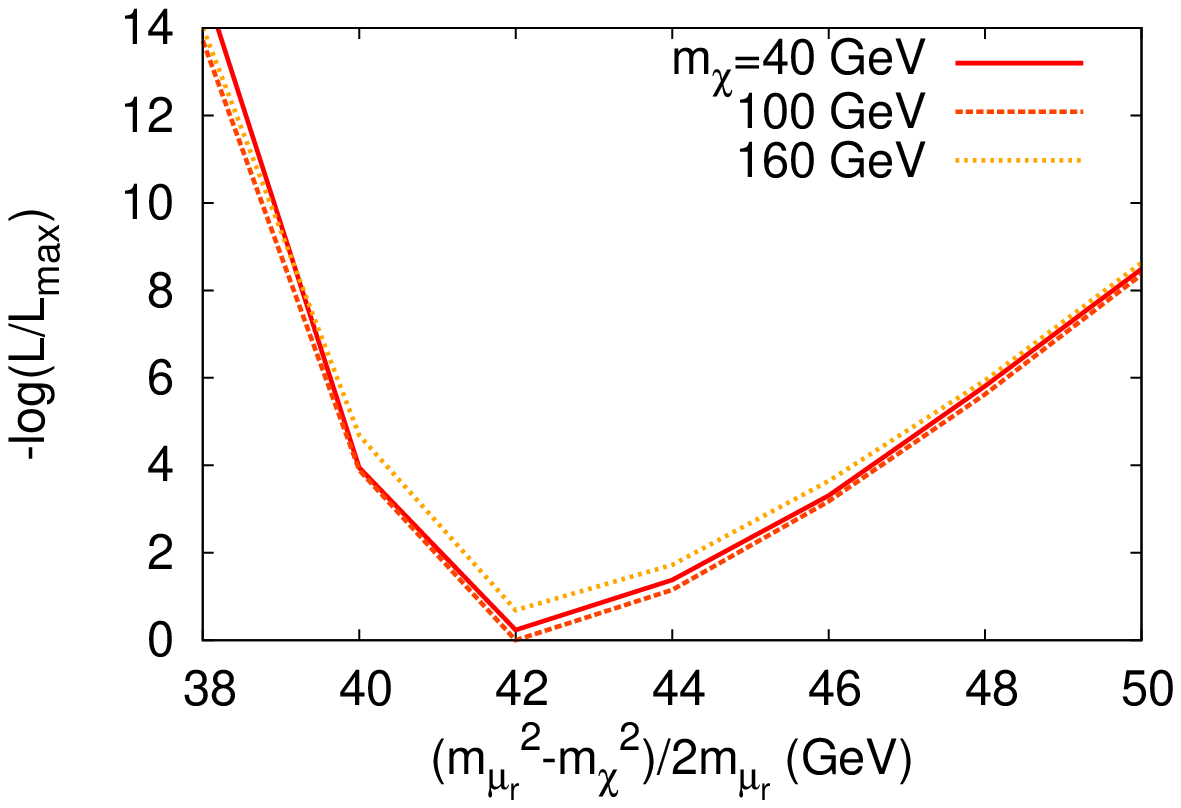}}
\subfigure[]{\includegraphics[scale=0.58]{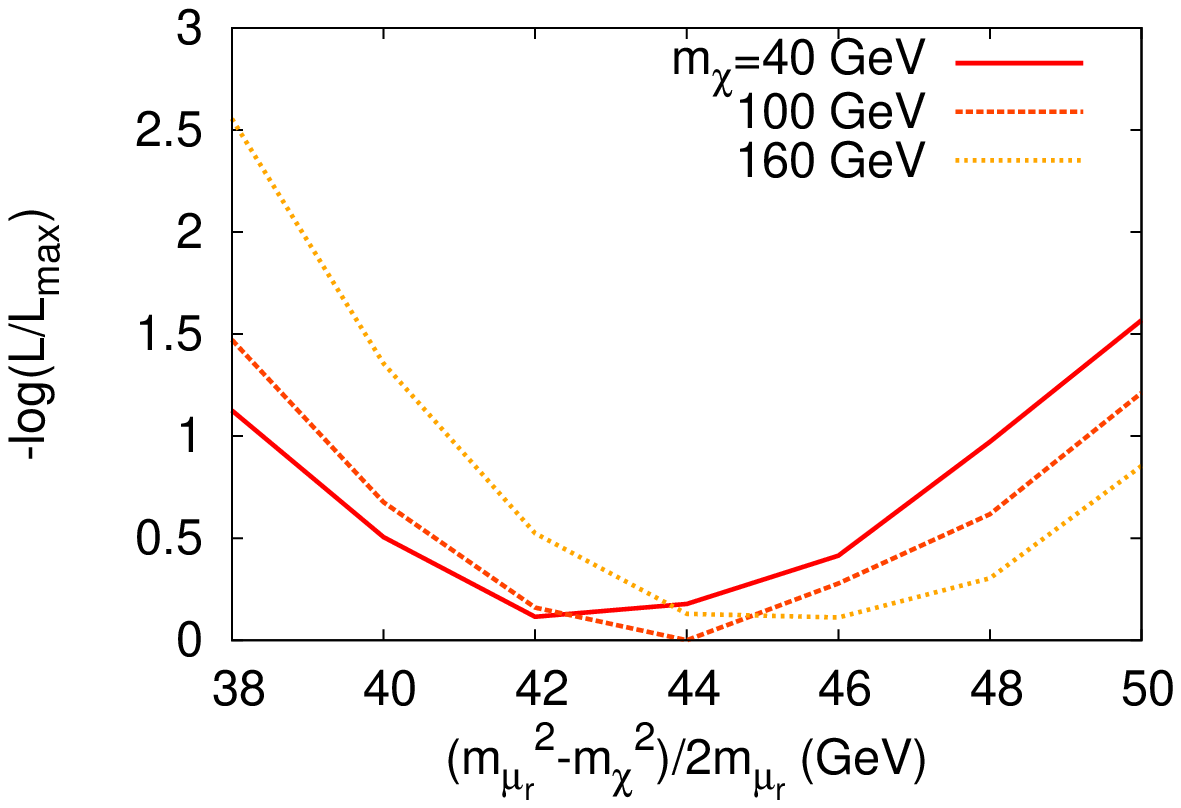}}
\caption{Logarithmic likelihood as a function of the hypothesis values for [$(m_{\tilde{\mu}_r}^2-m_{\tilde{\chi}_1}^2)/2m_{\tilde{\mu}_r},m_{\tilde{\chi}_1}$] built upon (a) the matrix element weights, (b) the transverse momentum of the $\mu^+$ and the invariant mass of the muons. }
\label{fig:2D_likelihood}
\end{figure}

We have considered the following input hypothesis for the masses of the sparticles:
\begin{equation}
m_{\tilde{\mu}_r, \textrm{input}}=150 \textrm{ GeV}, \quad m_{\tilde{\chi}_1, \textrm{input}}=100 \textrm{ GeV} \, .
\end{equation}
Under this hypothesis, we have generated events corresponding to the decay chain in Eq.~(\ref{susyDC}). We have built a sample of fifty events with exactly one $\mu^+$ and one $\mu^-$ with a transverse momentum larger than $20$ GeV, and no other particles except maybe some jets with a $p_T$ less than $20$ GeV. These events are regarded as a pseudo-experiment sample in the following.

The sensitivity that can be achieved with the matrix element method has been analysed by computing the weights $P(x_i|m_{\tilde{\mu}_r},m_{\tilde{\chi}_1})$ for each event $x_i$ in the pseudo-experiment sample. 
A bias may be introduced by the acceptance cuts. This effect has been corrected 
by normalizing the probability density in the acceptance region.
This amounts to replace the factor $1/\sigma_\alpha$ by the factor
$1/\sigma^{\textrm{obs}}_\alpha$ in the definition of the 
probability density in Eq~(\ref{weight_def}), with $\sigma^{\textrm{obs}}$ the 
cross section in the acceptance region.
In terms of the weights normalized in the acceptance region,
the unbiased likelihood has the usual form
\begin{equation}
\label{eq:likelidood_with_acc_term}
\log L(m_{\tilde{\mu}_r},m_{\tilde{\chi}_1})=\sum_{i=1}^{N=50} \log P(x_i|m_{\tilde{\mu}_r},m_{\tilde{\chi}_1}).
\end{equation}

It is advantageous to express the likelihood in term of the variable $(m_{\tilde{\mu}_r}^2-m_{\tilde{\chi}_1}^2)/2m_{\tilde{\mu}_r}$, that corresponds to the momentum of each final state particle in the rest frame of the smuon from which it originates. The complementary variable can be chosen to be $m_{\tilde{\chi}_1}$. The likelihood for different theoretical hypotheses is shown in Figure~\ref{fig:2D_likelihood}(a). The optimal value for the variable $(m_{\tilde{\mu}_r}^2-m_{\tilde{\chi}_1}^2)/2m_{\tilde{\mu}_r}$ is $42$ GeV, which corresponds to the input value. There is a very mild sensitivity with respect to variation of the complementary variable.

One way to highlight the increase of sensitivity by using the complete theoretical and experimental information is to compare the profile of the likelihood built upon the matrix element weights (shown in Figure~\ref{fig:2D_likelihood}(a)) with the likelihood profile that is obtained by keeping only the information contained in the transverse momentum $p_{T\mu}$ of the $\mu^+$ and the invariant mass $M_{\mu\mu}$ of the muons. In order to simplify the computation of this second likelihood profile we neglect the correlations  between the two variables $p_{T\mu}$ and $M_{\mu\mu}$. Thus the weight attached to each event is reduced to
\begin{equation}
P(x_i|m_{\tilde{\mu}_r},m_{\tilde{\chi}_1}) \rightarrow \frac{1}{\sigma} \frac{d\sigma}{dp_{T \mu}}(p_{T\mu}|m_{\tilde{\mu}_r},m_{\tilde{\chi}_1})
\times \frac{1}{\sigma} \frac{d\sigma}{d M^{\mu\mu}} (M_{\mu\mu}|m_{\tilde{\mu}_r},m_{\tilde{\chi_1}})
\end{equation}
The resulting likelihood profile is displayed in Figure~\ref{fig:2D_likelihood}(b). The comparison with Figure~\ref{fig:2D_likelihood}(a) shows that the sensitivity to the theoretical hypothesis  $(m_{\tilde{\mu}_r},m_{\tilde{\chi}_1})$ is dramatically reduced when only the information contained in the kinematic variables $p_{T\mu}$ and $M_{\mu\mu}$ is used.

\section{Conclusion}
\label{conclusion}

We have presented a new algorithm that allows the automatic
computation of the weights appearing in
the matrix element method. Given an arbitrary decay chain and a
transfer function tuned
to the resolution of the detector, our code produces a specific phase-space generator that
combines different phase-space mappings optimized for the integration
of the product of
the matrix element and the transfer function. The mappings are obtained
by applying a specific transformation to
 the standard
parametrization of the phase-space measure. 
This transformation is expressed in terms of a composition
of changes of variables acting on different kinematic sectors of the topology. 
As a result, our algorithm leads to a modular structure, and hence is particularly
convenient for
future improvements. For example, the current implementation could be
easily extended to
include non-analytical changes of  variables.

The availability of a tool such as MadWeight that  provides  the
resource for the automatic
evaluation of the weights, sparing the user to focus on the technical
details of matrix element
generation and integration over phase space, paves the way to a
potentially large number
of new applications.
First MadWeight could  be used to improve our understanding of the
matrix element method
itself and of its limitations. For example, its implementation within
Madgraph
provides all the required computational tools to analyse the influence
of additional jet
radiation in a specific measurement, or to estimate the systematic
error resulting from the
parametrization of the transfer function.
Second not only  measurement of masses or cross sections in Standard
Model could be achieved in
a effortless and more efficient way,  but the matrix element method
could also be employed in
the search and identification of new physics models. Different
hypotheses, such as those corresponding
to (any) new physics scenario and/or benchmark parameter points could
be tested against data and be
assigned a meaningful relative probability.
We look forward to exciting new developments in these directions.

\section{Acknowledgment} 
We are in debt with Tim Stelzer and Tilman Plehn  for motivating us and sharing  their enthusiasm on the matrix element method. 
We acknowledge many interesting discussions on this and related topics with Johan Alwall, Bob McElrath and the participants
of several very stimulating workshops on new physics searches, such as the FOCUS weeks organized in IPMU (Tokyo, Japan)
and in Aspen 2009. We thank the CP3 IT Team for all their help, advice and management of the cluster.
This work is partially funded by the HEPTOOLS EU network trough Marie Curie programme RTN MRTN-CT-
2006-035505, by Belgian Technical and Cultural Affairs through the Interuniversity Attraction Pole P6/11,
and by the Department of Energy (USA) under grant {DE-FG02-91-ER40690}.

\appendix
\section{Phase-space measure associated with the main blocks}
\label{appendix:MB}
\subsection{MB A}
\begin{figure}[here]
\center
\resizebox{!}{4cm}{ \includegraphics[]{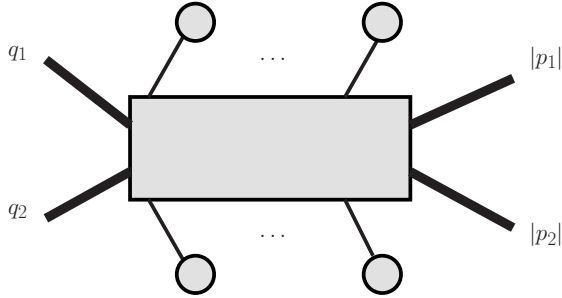}}
\caption{Notation for the kinematics of MB A.}
\label{AppMB_A}
\end{figure}

The notation for the phase-space variables
associated with this main block is given in Figure~\ref{AppMB_A}.
The two momenta $p_1$ and $p_2$ correspond to the visible particles that enter into the main block,
along with the Bjorken fractions $q_1$ and $q_2$. 
The standard phase-space parametrization associated with this main block reads
\begin{equation}
dq_1 dq_2 \frac{d^3p_1}{(2 \pi)^3 2 E_1}  \frac{d^3p_2}{(2 \pi)^3 2 E_2} (2 \pi)^4 \delta^4\left(P_{\textrm{in}}-P_{\textrm{fin}}\right) .
\end{equation} 
The four-vectors $P_{\textrm{in}}$ and $P_{\textrm{fin}}$ 
refer to the total momenta in the initial  and  final states, respectively.
In our procedure we apply a change of variables that leads to the following parametrization of the phase-space measure
\begin{equation}
\frac{1}{16 \pi^2 E_1 E_2} d \theta_1 d \phi_1  d \theta_2 d \phi_2 \times  J,
\end{equation}
where $\theta_i$ and $\phi_i$ refer to the polar and azimuthal angles of particle $i$ with respect to the beam axis.
The Jacobian $J$ of this transformation reads
\begin{equation} 
J=\frac{2}{s}|\bs p_1|^2 |\bs p_2|^2   |\cos \phi_1 \sin \phi_2 -\sin \phi_1 \cos \phi_2|^{-1} ,
\end{equation} 
where $s$ is the squared invariant mass of the colliding hadrons. 
The energies $E_1$, $E_2$ of the final particles in the main block are adjusted to balance the transverse momentum 
$\bs p_T^{\textrm{branches}}$
of all the branches represented by the blobs in Figure~\ref{AppMB_A}. We assume that this transverse 
momentum is different from zero, except maybe in a region of null measure 
(see the discussion in Section~\ref{sec:checks}). This requires the number of 
particles in the final states to be larger than or equal to three. Then
the variables $|\bs p_1|$, $|\bs p_2|$ can be expressed as the solution of the following linear system
\begin{subequations}
\begin{eqnarray} 
|\bs p_1| \sin \theta_1 \cos \phi_1 +|\bs p_2| \sin \theta_2 \cos \phi_2 &=& - p_{x}^{\textrm{branches}} 
\label{MB__A_Eqlin1} ,\\
|\bs p_1| \sin \theta_1 \sin \phi_1 +|\bs p_2| \sin \theta_2 \sin \phi_2 &=& -  p_{y}^{\textrm{branches}} . \label{MB__A_Eqlin2}
\end{eqnarray}
\end{subequations}
The Bjorken fractions $q_1,q_2$ are then fixed by imposing the conservation of  total energy and total momentum along the beam axis. 

\subsection{MB B}
\begin{figure}[here]
\center
\resizebox{!}{4cm}{ \includegraphics[]{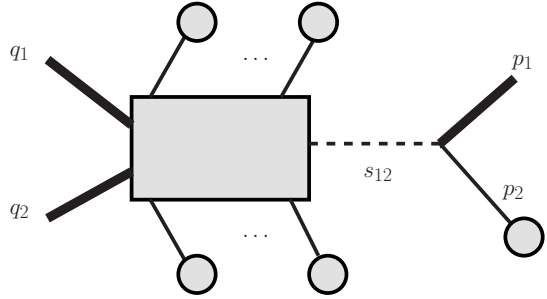}}
\caption{Notation for the kinematics of MB B.} 
\label{AppMB_B}
\end{figure}

The notation for the phase-space variables
associated with this main block is given in Figure~\ref{AppMB_B}.
The momentum $p_1$ corresponds to the missing particle  that belongs to the main block,
along with the Bjorken fractions $q_1$ and $q_2$. 
The momentum $p_2$ corresponds to the branch that is directly connected
to that missing particle. The variable $s_{12}$ is the invariant 
$(p_1+p_2)^2$. 
The standard phase-space parametrization associated with this constrained sector reads
\begin{equation}
dq_1 dq_2 \frac{d^3p_1}{(2 \pi)^3 2 E_1}  (2 \pi)^4 \delta^4\left(P_{\textrm{in}}-P_{\textrm{fin}}\right) .
\end{equation}
The four-vectors $P_{\textrm{in}}$ and $P_{\textrm{fin}}$
refer to the total momenta in the initial and final states, respectively.
In our procedure we apply a change of variables that leads to the following parametrization of the phase-space measure
\begin{equation}
\frac{1}{4 \pi  E_1 } d s_{12} \times  J .
\end{equation}
The Jacobian $J$ of this transformation is given by
\begin{equation}
J=\frac{E_1}{s}  |p_{2z}E_1-E_2p_{1z}|^{-1} ,
\label{jacMB_B}
\end{equation}
where $s$ is the squared invariant mass of the colliding hadrons.
The transverse momentum the missing particle is fixed by requiring 
that it balances the transverse momentum of all the branches represented by the blobs in Figure~\ref{AppMB_B}. 
The component $p_{1z}$ of momentum along the beam axis is fixed by imposing the 
invariant mass condition
\begin{equation}
(p_1+p_2)^2=s_{12} .
\label{condMB_B}
\end{equation}
If the energy $E_1$ of the missing particle is treated as an independent parameter, the left
side of Eq. (\ref{condMB_B}) is a first-order polynomial in $p_{1z}$. We therefore obtain a unique
expression for $p_{1z}$ in terms of $E_1$. The mass-shell condition associated with the missing particle
gives rise to up to two solutions for the energy $E_1$. Each solution that gives
a real positive value for $E_1$ and that leads to values of the Bjorken fractions $q_1$ and $q_2$
between 0 and 1
is  kept, as it corresponds to a distinct physical phase-space point at which the Jacobian in Eq.~(\ref{jacMB_B})
and the integrand must be evaluated.

\subsection{MB C}
\begin{figure}[here]
\center
\resizebox{!}{4cm}{ \includegraphics[]{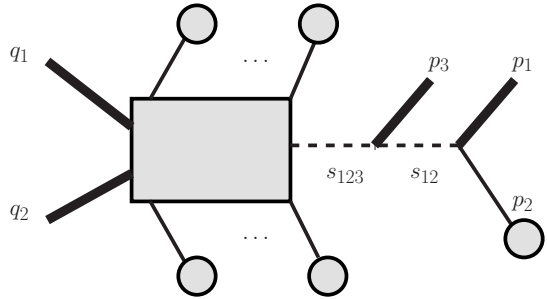}}
\caption{Notation for the kinematics of MB C.} 
\label{AppMB_C}
\end{figure}

The notation for the phase-space variables 
associated with this main blob is given in Figure~\ref{AppMB_C}.
The momentum of the missing particle is denoted by $p_1$, the momentum of the branch directly connected
to the missing particle is denoted by $p_2$, the momentum of the massless visible particle in the main block
is denoted by $p_3$.  The Bjorken fractions are denoted by $q_1$ and $q_2$.  The variables $s_{12}$ and $s_{123}$ refer to the invariants $(p_1+p_2)^2$
and $(p_1+p_2+p_3)^2$, respectively. 
The standard phase-space parametrization associated with this main block reads
\begin{equation}
dq_1 dq_2 \frac{d^3p_1}{(2 \pi)^3 2 E_1} \frac{d^3p_3}{(2 \pi)^3 2 E_3}  (2 \pi)^4 \delta^4\left(P_{\textrm{in}}-P_{\textrm{fin}}\right) .
\end{equation}
The four-vectors $P_{\textrm{in}}$ and $P_{\textrm{fin}}$
refer to the total momenta in the initial and final states, respectively.
In our procedure we apply a change of variables  that leads to the following parametrization of the phase-space measure
\begin{equation}
\frac{1}{16 \pi^2  E_1 E_3 } d\phi_3 d\theta_3 ds_{12} d s_{123} \times  J .
\end{equation}
The Jacobian $J$ of this transformation is given by
\begin{eqnarray}
J &=&  \sin \theta_3 \frac{E_3^2 E_1}{s} \bigg|  \chi E_2 p_{1z} -   \chi E_1 p_{2z} -  \nonumber  \\
  & &      2\cos(\phi_3)\cos(\theta_3)E_2 p_{1x}E_3\sin(\theta_3) +
        2\cos(\phi_3)\cos(\theta_3)E_1 p_{2x}E_3\sin(\theta_3) - \nonumber  \\
  & &      2\cos(\phi_3) p_{1z} p_{2x} E_3 \sin(\theta_3) +
        2\cos(\phi_3) p_{1x} p_{2z} E_3 \sin(\theta_3) - \nonumber  \\
  & &      2\cos(\theta_3) E_2 p_{1y} E_3 \sin(\phi_3)\sin(\theta_3) +
        2\cos(\theta_3) E_1 p_{2y} E_3 \sin(\phi_3)\sin(\theta_3) -  \nonumber  \\
  & &      2 p_{1z} p_{2y} E_3\sin(\phi_3)\sin(\theta_3) +
        2 p_{1y} p_{2z} E_3 \sin(\phi_3) \sin(\theta_3) + \nonumber  \\
  & &      2\cos(\phi_3)^2 E_2 p_{1z} E_3 \sin(\theta_3)^2 -
        2\cos(\phi_3)^2 E_1 p_{2z} E_3 \sin(\theta_3)^2 + \nonumber  \\
  & &      2 E_2 p_{1z} E_3 \sin(\phi_3)^2 \sin(\theta_3)^2 -
        2 E_1 p_{2z} E_3 \sin(\phi_3)^2 \sin(\theta_3)^2  \bigg|^{-1} ,
        \label{jacMB_C}
\end{eqnarray}
with $\chi=2p_3.(p_1+p_2)/E_3$ and $s$ standing for the
squared invariant mass of the colliding hadrons.

If we treat the variables $E_1$ and $\alpha=2p_1.p_3$ as two independent parameters, 
the components of the three-momentum $\bs p_1$ of the missing particle 
and the energy $E_3=|\bs p_3|$ of the massless visible particle can be expressed as the solution
of the following linear system of four equations
\begin{subequations}
\begin{eqnarray}
(p_1+p_2)^2 & = & s_{12} \\
(p_1+p_2+p_3)^2 & = & s_{123} \\
p_{1x}+E_3 \sin \theta_3 \cos \phi_3 &=& -p_{Tx}^{\textrm{branches}} \\
p_{1y}+E_3 \sin \theta_3 \sin \phi_3 &= & -p_{Ty}^{\textrm{branches}} 
\end{eqnarray}
\end{subequations}
that is parametrized by the momentum $p_2$, by the angles $ \theta_3$ and $\phi_3$, by the total transverse
momentum $\bs p_T^{\textrm{branches}}$ of all the branches represented, by the blobs in Figure~\ref{AppMB_C} and by the 
variables $\alpha$ and $E_1$. The next step is to determine the values of the variables 
$\alpha$ and $E_1$. The mass-shell condition for the missing particle of momentum $p_1$ and  
the equation $2p_1.p_3=\alpha$ defines a system of two coupled quadratic equations
in the variables $E_1$ and $\alpha$, parametrized by the momenta of the blocks. 
This system can be solved analytically. There are up to
four solutions for $E_1$ and $\alpha$. Each solution that is physical ({\it i.e.}, such that
$|p_3|>0, \, E_1>0$ and each of the Bjorken fractions $q_1,q_2$ is between $0$ and $1$)    
corresponds to a distinct  phase-space point  at which the Jacobian in Eq.~(\ref{jacMB_C})
and the integrand must be evaluated.

\subsection{MB D}
\begin{figure}[here]\center
\resizebox{!}{4cm}{ \includegraphics[]{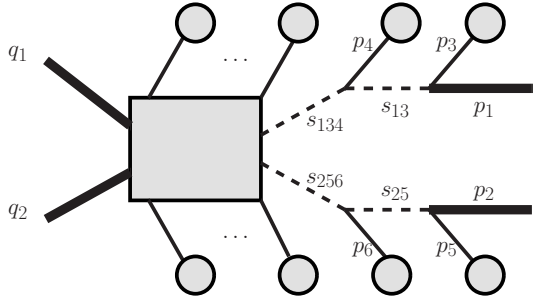}}\caption{Notation for the kinematics of MB D.}
\label{AppMB_D}\end{figure}

The notation for the phase-space variables
associated with this constrained sector is given in Figure~\ref{AppMB_D}.
The momenta of the missing particles are denoted by $p_1$ and $p_2$, the momenta of the branches 
connected to the main block are denoted by $p_3, \, p_4, \, p_5$ and $p_6$.   The Bjorken fractions are denoted by $q_1$ and $q_2$. 
The variables $s_{ij}$ and $s_{ijk}$ refer to the invariants $(p_i+p_j)^2$
and $(p_i+p_j+p_k)^2$, respectively. The standard phase-space parametrization associated with this main block reads
\begin{equation}
dq_1 dq_2 \frac{d^3p_1}{(2 \pi)^3 2 E_1} \frac{d^3p_2}{(2 \pi)^3 2 E_2} 
 (2 \pi)^4 \delta^4\left(P_{\textrm{in}}-P_{\textrm{fin}}\right) .
\end{equation}
The four-vectors $P_{\textrm{in}}$ and $P_{\textrm{fin}}$
refer to the total momenta in the initial and final states,
 respectively.
In our procedure we apply a change of variables that leads to the following parametrization of the phase-space measure
\begin{equation}
\frac{1}{16 \pi^2  E_1 E_2 } ds_{13} ds_{134} d s_{25} d s_{256} \times  J .
\end{equation}
The Jacobian $J$ of this transformation is given by
\begin{eqnarray}
J & = & \frac{E_1 E_2}{8s}  \bigg| E_3 \Big\{ E_5 \big[ 
              p_{34z} (p_{1y}p_{2z}p_{56x} - p_{1x}p_{2z}p_{56y} \nonumber \\ & &  - p_{1y} p_{2x} p_{56z} +
                   p_{1x}p_{2y}p_{56z}) + 
                p_{1z}(-p_{2z}p_{34y}p_{56x} +  \nonumber \\ & &  p_{2z}p_{34x}p_{56y} -     
                   p_{2y}p_{34x}p_{56z} + p_{2x}p_{34y}p_{56z})\big] + \nonumber \\ & &
             (E_{56}p_{2z} - E_{2}p_{56z})
              (p_{1z}p_{34y}p_{5x} - p_{1y}p_{34z}p_{5x} - p_{1z}p_{34x}p_{5y} + \nonumber \\ & &
                p_{1x}p_{34z}p_{5y}) +
             \big[ E_{56}(p_{1z}p_{2y}p_{34x} - p_{1z}p_{2x}p_{34y} + p_{1y}p_{2x}p_{34z} - \nonumber \\ & &
                   p_{1x}p_{2y}p_{34z}) +
                E_2(p_{1z}p_{34y}p_{56x} - p_{1y}p_{34z}p_{56x} - p_{1z}p_{34x}p_{56y} + \nonumber \\ & &
                   p_{1x}p_{34z}p_{56y})\big] p_{5z} \Big\} +
          E_{34} \Big\{ E_5 p_{2z} (p_{1z} p_{3y} p_{56x} - p_{1y} p_{3z} p_{56x} \nonumber \\ & & - p_{1z} p_{3x} p_{56y} + 
                p_{1x} p_{3z} p_{56y}) +
             E_5 (p_{1z} p_{2y} p_{3x}   - p_{1z} p_{2x} p_{3y} \nonumber \\ & & + p_{1y} p_{2x} p_{3z} - 
                p_{1x} p_{2y} p_{3z}) p_{56z} -
             ( E_{56} p_{2z} - E_2 p_{56z})  \nonumber \\ & &
              (p_{1z} p_{3y} p_{5x} - p_{1y} p_{3z} p_{5x} - p_{1z} p_{3x} p_{5y} + p_{1x} p_{3z} p_{5y}) \nonumber \\ & &
               - \big[ E_{56} (p_{1z} p_{2y} p_{3x} - p_{1z} p_{2x} p_{3y} + p_{1y} p_{2x} p_{3z} -
                   p_{1x} p_{2y} p_{3z}) +\nonumber  \\ & &
                E_2 (p_{1z} p_{3y} p_{56x} - p_{1y} p_{3z} p_{56x} - p_{1z} p_{3x} p_{56y} +
                   p_{1x} p_{3z} p_{56y})\big] p_{5z} \Big\} +\nonumber \\ & &
          E_1 \Big\{ \big[ E_5 (p_{2z} (-p_{34z} p_{3y} p_{56x} + p_{34y} p_{3z} p_{56x} +\nonumber \\ & &
                   p_{34z} p_{3x} p_{56y} - p_{34x} p_{3z} p_{56y}) +\nonumber \\ & &
                (-p_{2y} p_{34z} p_{3x} + p_{2x} p_{34z} p_{3y} + p_{2y} p_{34x} p_{3z} -
                   p_{2x} p_{34y} p_{3z}) p_{56z}\big] +\nonumber \\ & &
             \big[ E_{56} p_{2z} - E_2 p_{56z}) 
              (p_{34z} p_{3y} p_{5x} - p_{34y} p_{3z} p_{5x} - p_{34z} p_{3x} p_{5y} +\nonumber \\ & &
                p_{34x} p_{3z} p_{5y}) +
             ( E_{56} (p_{2y} p_{34z} p_{3x} - p_{2x} p_{34z} p_{3y} - p_{2y} p_{34x} p_{3z} +\nonumber \\ & &
                   p_{2x} p_{34y} p_{3z}) +
                E_2 (p_{34z} p_{3y} p_{56x} - p_{34y} p_{3z} p_{56x} - p_{34z} p_{3x} p_{56y} +\nonumber \\ & &
                   p_{34x} p_{3z} p_{56y}) \big] p_{5z} \Big\}  \bigg|^{-1} ,
\label{jacMB_D}
\end{eqnarray}
where $E_{ij}=E_i+E_j$,  $p_{ij}=p_i+p_j$, and $s$ is the
squared invariant mass of the colliding hadrons in their center-of-mass frame.
 If we treat the variables $E_1$ and $E_2$ as independent parameters, then
the components of the three-momenta $\bs{p}_1, \bs{p}_2$ of the missing particles 
 can be expressed as the solution
of the following linear system of six equations
\begin{subequations}
\begin{eqnarray}
(p_1+p_3)^2 & = & s_{13} \\
(p_1+p_3+p_4)^2 & = & s_{134} \\
(p_2+p_5)^2 & = & s_{25} \\
(p_2+p_5+p_6)^2 & = & s_{256} \\
p_{1x}+p_{2x} &=& -p_{Tx}^{\textrm{branches}} \\
p_{1y} +p_{2y}&= & -p_{Ty}^{\textrm{branches}} 
\end{eqnarray}
\end{subequations}
that is  parametrized by the momenta $p_3, \dots,p_6$ of the branches connected to the main block, by 
the total transverse momentum $\bs p_{T}^{\textrm{branches}}$ of all the branches 
represented by the 
blobs in Figure~\ref{AppMB_D} and by the 
variables $E_1$ and $E_2$. The next step is to determine the values of the variables 
$E_1$ and $E_2$. The mass-shell conditions for the two missing particles of momentum 
$p_1$ and $p_2$ define a system of two coupled quadratic equations
in the variables $E_1$ and $E_2$, that can be solved analytically. There are up to
four solutions for $E_1$ and $E_2$. Each solution that is physical ({\it i.e.}, such that
$E_2>0, \, E_1>0$ and each of the Bjorken fractions $q_1,q_2$ is between $0$ and $1$)    
corresponds to a distinct  phase-space point at which 
 the Jacobian in Eq.~(\ref{jacMB_D})
and the integrand must be evaluated.

 \subsection{MB E}
\begin{figure}[here]\center
\resizebox{!}{4cm}{ \includegraphics[]{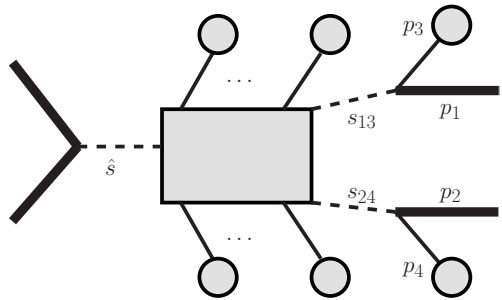}}\caption{Notation for the kinematics of MB E.}
\label{AppMB_E}\end{figure}

 The notation for the phase-space variables
associated with this main blob is given in Figure~\ref{AppMB_E}.
The momenta of the missing particles are denoted by $p_1$ and $p_2$, the momenta of the branches directly 
connected to these missing particles are denoted by $p_3$ and $p_4$.  The Bjorken fractions are denoted by $q_1$ and $q_2$.
The variables $s_{ij}$ refer to the invariants $(p_i+p_j)^2$ and $\hat s$ denotes the 
squared invariant mass of the colliding partons. 
 The standard phase-space parametrization associated with this MB reads
\begin{equation}
dq_1 dq_2 \frac{d^3p_1}{(2 \pi)^3 2 E_1} \frac{d^3p_2}{(2 \pi)^3 2 E_2} 
 (2 \pi)^4 \delta^4\left(P_{\textrm{in}}-P_{\textrm{fin}}\right) .
\end{equation}
The four-vectors $P_{\textrm{in}}$ and $P_{\textrm{fin}}$
refer to the total momenta in the initial and final states,
 respectively.
In our procedure we apply a change of variables that leads to the following parametrization of the phase-space measure
\begin{equation}
\frac{1}{16 \pi^2  E_1 E_2 } dy d\hat{s} ds_{13} d s_{24}  \times  J ,
\end{equation}
where $y$ is the rapidity of the colliding partons in the lab frame.  
The Jacobian $J$ of this transformation is given by
\begin{eqnarray}
J= \frac{E_1 E_2}{4s} \bigg|  E_4 ( p_{1z}  p_{2y}  p_{3x} -  p_{1y}  p_{2z}  p_{3x}    -  p_{1z}  p_{2x}  p_{3y} + 
             p_{1x}  p_{2z}  p_{3y} +  p_{1y}  p_{2x}  p_{3z} -   \nonumber & & \\ p_{1x}  p_{2y}  p_{3z}) + 
           E_2  p_{1z}  p_{3y}  p_{4x} -  E_1  p_{2z}  p_{3y}  p_{4x} -  E_2  p_{1y}  p_{3z}  p_{4x} + 
           E_1  p_{2y}  p_{3z}  p_{4x}    \nonumber & & \\ -  E_2  p_{1z}  p_{3x}  p_{4y} +  E_1  p_{2z}  p_{3x}  p_{4y} +
           E_2  p_{1x}  p_{3z}  p_{4y} -  E_1  p_{2x}  p_{3z}  p_{4y} + 
          ( E_2  p_{1y}  p_{3x} +  \nonumber & & \\  -  E_1  p_{2y}  p_{3x} -  E_2  p_{1x}  p_{3y} +  E_1  p_{2x}  p_{3y})  p_{4z}  +
     E_3 (- p_{1z}  p_{2y}  p_{4x} +  p_{1y}  p_{2z}  p_{4x}  \nonumber & & \\ +  p_{1z}  p_{2x}  p_{4y} -
              p_{1x}  p_{2z}  p_{4y} -  p_{1y}  p_{2x}  p_{4z} +  p_{1x}  p_{2y}  p_{4z}) \bigg|^{-1}  
              \label{jacMB_E} .
\end{eqnarray} 
 
 If we treat the variables $E_1$, $E_2$ and $p_{2y}$ as independent parameters, 
the other components of the momenta $ p_1,  p_2$ of the missing particles 
 can be expressed as the solution
of the following linear system of five equations
\begin{subequations}
\begin{eqnarray}
(p_1+p_3)^2 & = & s_{13} \\
(p_2+p_4)^2 & = & s_{24} \\
p_{1x}+p_{2x} &=& -p_{x}^{\textrm{branches}} \\
p_{1y} +p_{2y}&= & -p_{y}^{\textrm{branches}} \\
p_{1z}+p_{2z}& = & \sinh (y)  \hat {s}^{1/2}-p_{z}^{\textrm{branches}}  
\end{eqnarray}
\end{subequations}
that is parametrized by the momenta of the branches $p_3$ and $p_4$, by the total
momentum $p^{\textrm{branches}}$ of all the branches represented by the blobs in Figure~\ref{AppMB_E},
by the rapidity $y$ and the invariant mass $\hat {s}^{1/2}$ of the colliding partons and by the 
variables $E_1$, $E_2$ and $p_{2y}$. The next step is to fix the values of the variables 
$E_1$, $E_2$ and $p_{2y}$. The variable $E_1$ can be expressed as a linear function of $E_2$:
\begin{equation}
E_1  =  \cosh (y) \hat {s}^{1/2}-E^{\textrm{branches}} -E_2 .
\end{equation}
Then the mass-shell conditions for the two missing particles define a system of two coupled quadratic equations
in the variables $E_2$ and $p_{2y}$. In this case, the quartic terms of the two equations have the same coefficients, and the system
reduces to a linear equation and a quadratic equation. There are up to
two solutions for $E_2$ and $p_{2y}$. Each solution that is physical ({\it i.e.}, such that
$E_2>0, \, E_1>0$)    
corresponds to a distinct  phase-space point
at which 
 the Jacobian in Eq.~(\ref{jacMB_E})
and the integrand must be evaluated.

  \subsection{MB F}
\begin{figure}[here]\center
\resizebox{!}{4cm}{ \includegraphics[]{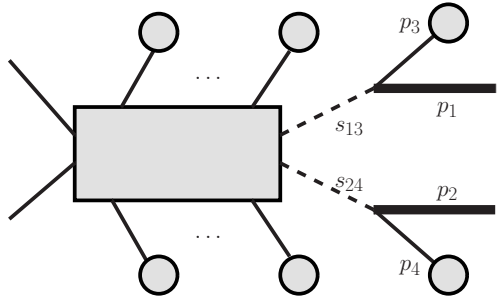}}\caption{Notation for the kinematics of MB F.}
\label{AppMB_F}\end{figure}

 The notation for the phase-space variables
associated with this  constrained sector is given in Figure~\ref{AppMB_F}.
The momenta of the missing particles are denoted by $p_1$ and $p_2$, the momenta of the branches 
directly connected to these missing particles are denoted by $p_3$ and $p_4$.  The Bjorken fractions are denoted by $q_1$ and $q_2$.
The variables $s_{ij}$ refer to the invariants $(p_i+p_j)^2$.
 The standard phase-space parametrization associated with this main block reads
\begin{equation}
dq_1 dq_2 \frac{d^3p_1}{(2 \pi)^3 2 E_1} \frac{d^3p_2}{(2 \pi)^3 2 E_2} 
 (2 \pi)^4 \delta^4\left(P_{\textrm{in}}-P_{\textrm{fin}}\right) .
\end{equation}
The four-vectors $P_{\textrm{in}}$ and $P_{\textrm{fin}}$
refer to the total momenta in the initial and final states,
 respectively.
In our procedure we apply a change of variables that leads to the following parametrization of the phase-space measure
\begin{equation}
\frac{1}{16 \pi^2  E_1 E_2 } dq_1 dq_2 ds_{13} d s_{24}  \times  J .
\end{equation}
The Jacobian $J$ of this transformation   is given by
\begin{eqnarray}
J= \frac{E_1 E_2}{4} \bigg|   E_4 ( p_{1z}  p_{2y}  p_{3x} -  p_{1y}  p_{2z}  p_{3x}    -  p_{1z}  p_{2x}  p_{3y} + 
             p_{1x}  p_{2z}  p_{3y} +  p_{1y}  p_{2x}  p_{3z} -   \nonumber & & \\ p_{1x}  p_{2y}  p_{3z}) + 
           E_2  p_{1z}  p_{3y}  p_{4x} -  E_1  p_{2z}  p_{3y}  p_{4x} -  E_2  p_{1y}  p_{3z}  p_{4x} + 
           E_1  p_{2y}  p_{3z}  p_{4x}    \nonumber & & \\ -  E_2  p_{1z}  p_{3x}  p_{4y} +  E_1  p_{2z}  p_{3x}  p_{4y} +
           E_2  p_{1x}  p_{3z}  p_{4y} -  E_1  p_{2x}  p_{3z}  p_{4y} + 
          ( E_2  p_{1y}  p_{3x} +  \nonumber & & \\  -  E_1  p_{2y}  p_{3x} -  E_2  p_{1x}  p_{3y} +  E_1  p_{2x}  p_{3y})  p_{4z}  
    + E_3 (- p_{1z}  p_{2y}  p_{4x} +  p_{1y}  p_{2z}  p_{4x}  \nonumber & & \\ +  p_{1z}  p_{2x}  p_{4y} -
              p_{1x}  p_{2z}  p_{4y} -  p_{1y}  p_{2x}  p_{4z} +  p_{1x}  p_{2y}  p_{4z})  \bigg|^{-1}  .
              \label{jacMB_F}
\end{eqnarray} 
 
 If we treat the variables $E_1$, $E_2$ and $p_{2y}$ as independent parameters, 
the other components of the momenta $p_1, p_2$ of the missing particles 
 can be expressed as the solution
of the following linear system of five equations, 
\begin{subequations}
\begin{eqnarray}
(p_1+p_3)^2 & = & s_{13} \\
(p_2+p_4)^2 & = & s_{24} \\
p_{1x}+p_{2x} &=& -p_{x}^{\textrm{branches}} \\
p_{1y} +p_{2y}&= & -p_{y}^{\textrm{branches}} \\
p_{1z}+p_{2z} & = &  {s}^{1/2} (q_1-q_2)/2  -p_{z}^{\textrm{branches}}  
\end{eqnarray}
\end{subequations}
that is parametrized by the momenta $p_3$ and $p_4$, by  the total  
momentum $p^{\textrm{branches}}$
of all the branches represented by the blobs in Figure~\ref{AppMB_F},
by the Bjorken fractions $q_1$, $q_2$  and by the 
variables $E_1$, $E_2$ and $p_{2y}$. The next step is to fix the values of the variables 
$E_1$, $E_2$ and $p_{2y}$. 
The variable $E_1$ can be expressed as a linear function of $E_2$:
\begin{equation}
E_1  =  {s}^{1/2} (q_1+q_2)/2 -E^{\textrm{branches}} -E_2 .
\end{equation}
The mass-shell conditions for the two missing particles 
with momenta $p_1$ and $p_2$ define a system of two coupled quadratic equations
in the variables $E_2$ and $p_{2y}$. In this case, the quartic terms of the two equations have the same coefficients, and the system
reduces to a linear equation and a quadratic equation. There are up to
two solutions for $E_2$ and $p_{2y}$. Each solution that is physical ({\it i.e.}, such that
$E_2>0, \, E_1>0$)    
corresponds to a distinct  phase-space point
at which 
 the Jacobian in Eq.~(\ref{jacMB_F})
and the integrand must be evaluated.

\section{Phase-space measure associated with the secondary blocks}

\label{appendix:blocks}

\subsection{SB A}

\begin{figure}[here]\center
\resizebox{!}{3cm}{ \includegraphics[]{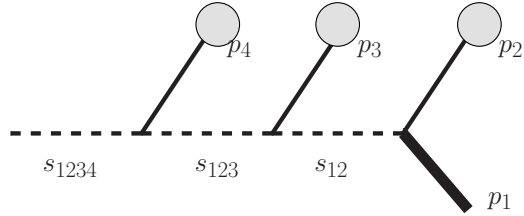}}\caption{Notation for the kinematics of SB A.}
\label{AppBLOCKA}\end{figure}

The notation for the phase-space variables
associated with this secondary block is given in Figure~\ref{AppBLOCKA}.
The momentum of the missing particle is denoted by $p_1$, the momenta of the three branches 
connected to the block are denoted by $p_2$,  $p_3$ and $p_4$.
The variables $s_{12}$, $s_{123}$ and $s_{1234}$  refer to the invariants $(p_1+p_2)^2$,
$(p_1+p_2+p_3)^2$, and $(p_1+p_2+p_3+p_4)^2$, respectively.
 The standard phase-space parametrization associated with this block reads
\begin{equation}
\frac{d^3p_1}{(2 \pi)^3 2 E_1}  .
\end{equation}
In our procedure we apply a change of variables that leads to the following parametrization of the phase-space measure
\begin{equation}
\frac{1}{(2 \pi)^3 2 E_1  } ds_{12} d s_{123}  d s_{1234}   \times  J .
\end{equation}
The Jacobian $J$ of this transformation   is given by
\begin{eqnarray}
J=\frac{E_1}{8} \Big| E_4(p_{1z}p_{2y}p_{3x} - p_{1y}p_{2z}p_{3x} -  
             p_{1z}p_{2x}p_{3y} + p_{1x}p_{2z}p_{3y} + p_{1y}p_{2x}p_{3z} & & \nonumber \\  - p_{1x}p_{2y}p_{3z}) 
           + E_{2}p_{1z}p_{3y}p_{4x} - E_{1}p_{2z}p_{3y}p_{4x} - E_2p_{1y}p_{3z}p_{4x} +
          E_1p_{2y}p_{3z}p_{4x}  & & \nonumber \\  - E_2p_{1z}p_{3x}p_{4y} + E_1p_{2z}p_{3x}p_{4y} + 
          E_2p_{1x}p_{3z}p_{4y} - E_1p_{2x}p_{3z}p_{4y} +
          (E_2p_{1y}p_{3x}  & & \nonumber \\  - E_1p_{2y}p_{3x} - E_2p_{1x}p_{3y} + E_1p_{2x}p_{3y})p_{4z} +
          E_3(-p_{1z}p_{2y}p_{4x} + p_{1y}p_{2z}p_{4x}   & & \nonumber \\ + p_{1z}p_{2x}p_{4y} -
             p_{1x}p_{2z}p_{4y} - p_{1y}p_{2x}p_{4z} + p_{1x}p_{2y}p_{4z}) \Big|^{-1} .
              \label{jacBLOCKA}
\end{eqnarray} 
  If we treat the variable $E_1$  as an independent parameter, 
the components of the three-momentum $\bs p_1$ of the missing particle
 can be expressed as the solution
of the following linear system of three equations
\begin{subequations}
\begin{eqnarray}
(p_1+p_2)^2 & = & s_{12} \\
(p_1+p_2 + p_3)^2 & = & s_{123} \\
(p_1+p_2 + p_3+p_4)^2 & = & s_{1234} 
\end{eqnarray}
\end{subequations}
that is parametrized by the momenta $p_2$ ,$p_3$, $p_4$ of the branches
connected to the block,  and by the 
variable $E_1$. The next step is to fix the value of the variable 
$E_1$. The mass-shell condition for the  missing particle
with momentum $p_1$ defines a quadratic equation
in the variable $E_1$. There are up to
two solutions for $E_1$. Each solution that is physical ({\it i.e.}, such that
$ E_1>0$)    
corresponds to a distinct  phase-space point
at which 
 the Jacobian in Eq.~(\ref{jacBLOCKA})
and the integrand must be evaluated.

\subsection{SB B}

\begin{figure}[here]
\center
\resizebox{!}{3cm}{ \includegraphics[]{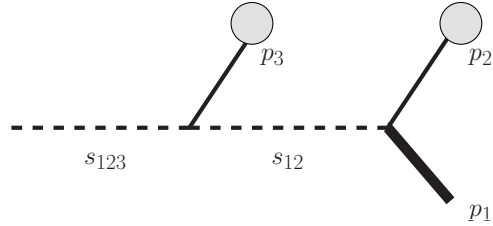}}\caption{Notation for the kinematics of SB B.}
\label{AppBLOCKB}\end{figure}

The notation for the phase-space variables
associated with this secondary block is given in Figure~\ref{AppBLOCKB}.
The momentum of the missing particle is denoted by $p_1$, the momenta of the two branches
connected to the block are denoted by $p_2$ and  $p_3$.
The variables $s_{12}$ and $s_{123}$  refer to the invariants $(p_1+p_2)^2$,
and $(p_1+p_2+p_3)^2$, respectively.
 The standard phase-space parametrization associated with this block reads
\begin{equation}
\frac{d^3p_1}{(2 \pi)^3 2 E_1} .
\end{equation}
In our procedure we apply a change of variables that leads to the following parametrization of the phase-space measure
\begin{equation}
\frac{1}{(2 \pi)^3 2 E_1  } d\phi_1 ds_{12} d s_{123}   \times  J ,
\end{equation}
where $\phi_i$ denotes the azimuthal angle of particle $i$.
The Jacobian $J$ of this transformation   is given by
\begin{eqnarray}
J= \frac{E_1}{4} p_{1T} \Big| -\cos (\phi_1-\phi_2) E_3 p_{2T} p_{1z}
      + \cos (\phi_1-\phi_3) E_2 p_{3T} p_{1z}   
      + E_3 p_{1T} p_{2z}  & & \nonumber \\  -  \cos(\phi_1-\phi_3) E_1 p_{3T} p_{2z}
      - E_2 p_{1T} p_{3z} +  \cos(\phi_1-\phi_2) E_1p_{2T} p_{3z}   \Big|^{-1}  . \label{jacBLOCKB}
\end{eqnarray} 
  If we treat the variable $E_1$  as an independent parameter, 
 the transverse momentum $ p_{1T}$ and the momentum component 
$p_{1z}$  of the missing particle
 can be expressed as the solution
of the following linear system of two equations
\begin{subequations}
\begin{eqnarray}
(p_1+p_2)^2 & = & s_{12} \\
(p_1+p_2 + p_3)^2 & = & s_{123} 
\end{eqnarray}
\end{subequations}
that is parametrized by the momenta $p_2$ and $p_3$ of the branches
connected to the block,  by the azimuthal angle $\phi_1$ and by the 
variable $E_1$. The next step is to fix the value of the variable 
$E_1$. The mass-shell condition for the  missing particle
with momentum $p_1$ defines a quadratic equation
in the variable $E_1$. There are up to
two solutions for $E_1$. Each solution that is physical ({\it i.e.}, such that
$ E_1>0$)    
corresponds to a distinct  phase-space point
at which 
 the Jacobian in Eq.~(\ref{jacBLOCKB})
and the integrand must be evaluated.

\subsection{SB C/D}

\begin{figure}[here]\center
\resizebox{!}{3cm}{ \includegraphics[]{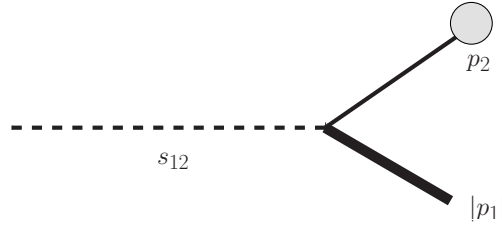}}\caption{Notation for the kinematics of SB C/D.}
\label{AppblockC}\end{figure}

The notation for the phase-space variables
associated with this secondary block is given in Figure~\ref{AppblockC}.
The momentum of the missing particle is denoted by $p_1$, the momentum of the  branch 
connected to the block  is denoted by $p_2$.
The variable $s_{12}$  refers to the invariant $(p_1+p_2)^2$.
 The standard phase-space parametrization associated with this block reads
\begin{equation}
\frac{d^3p_1}{(2 \pi)^3 2 E_1} .
\end{equation}
In our procedure we apply a change of variables that leads to the following parametrization of the phase-space measure
\begin{equation}
\frac{1}{(2 \pi)^3 2 E_1  } d\phi_1 d \theta_1 ds_{12}   \times  J ,
\end{equation}
where $\theta_1$ and $\phi_1$ denote the polar and azimuthal angles of the missing particle.
The Jacobian $J$ of this transformation   is given by
\begin{eqnarray}
J=\frac{E_1}{2} \sin \theta_1 |\bs p_1|^2 \bigg|  |\bs p_1| E_2- E_1 \hat {\bs {p}}_1.\bs p_2   \bigg|^{-1}          \label{jacBLOCKC} .
\end{eqnarray} 
  If we treat the variable $E_1$  as an independent parameter, 
the momentum modulus $|\bs p_1|$ of the missing particle
 can be expressed as the solution
of the following linear  equation
\begin{equation}
(p_1+p_2)^2  =  s_{12}
\end{equation}
that is parametrized by the momentum $p_2$ of the branch
connected to the block, by the polar and azimuthal angles 
$\theta_1$, $\phi_1$,  and by the 
variable $E_1$. The next step is to fix the value of the variable 
$E_1$. The mass-shell condition for the  missing particle
with momentum $p_1$ defines a quadratic equation
in the variable $E_1$. There are up to
two solutions for $E_1$. Each solution that is physical ({\it i.e.}, such that
$ E_1>0$)    
corresponds to a distinct  phase-space point
at which 
 the Jacobian in Eq.~(\ref{jacBLOCKC})
and the integrand must be evaluated.

\subsection{SB E}

\begin{figure}[here]\center 
\resizebox{!}{3cm}{ \includegraphics[]{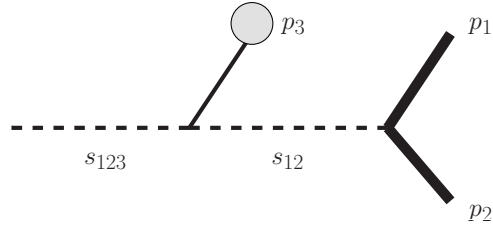}}\caption{Notation of the kinematics for SB E.}
\label{AppBLOCKD}\end{figure}

The notation for the phase-space variables
associated with this secondary block is given in Figure~\ref{AppBLOCKD}.
The momenta of the visible particles are denoted by 
 $p_1$ and $p_2$, the momentum of the branch
connected to the block  is denoted by $p_3$.
The variables $s_{12}$ and $s_{123}$  refer to the invariants $(p_1+p_2)^2$,
and $(p_1+p_2+p_3)^2$.
 The standard phase-space parametrization associated with this block reads
\begin{equation}
\frac{d^3p_1}{(2 \pi)^3 2 E_1}  \frac{d^3p_2}{(2 \pi)^3 2 E_2} .
\end{equation}
In our procedure we apply a change of variables that leads to the following parametrization of the phase-space measure
\begin{equation}
\frac{1}{(2 \pi)^6 4 E_1 E_2  } d \theta_1 d\phi_1 d \theta_2 d\phi_2 ds_{12} d s_{123}    \times  J ,
\end{equation}
where $\theta_i$ and $\phi_i$ denote the polar and azimuthal angles of particle $i$.
The Jacobian $J$ of this transformation   is given by
\begin{eqnarray}
J=\frac{E_2^2}{4} |\bs p_1|^2 \sin \theta_1 \sin \theta_2 \Big| (|\bs p_1||\bs p_2|/E_1-
|\bs p_2| f_{12} )(E_3-|\bs p_3| f_{23})  & & \nonumber \\ -(E_3 |\bs p_1|/E_1-
 |\bs p_3|f_{13})
(E_1-f_{12} |\bs p_1|)   \Big|^{-1}          \label{jacBLOCKD}
\end{eqnarray} 
where $f_{ij}$ stands for $\bs p_i. \bs p_j/|\bs p_i| |\bs p_j|$.
The values for the momenta $|\bs p_1|$ and  $|\bs p_2|$
can be obtained by solving the following linear system of equations
\begin{subequations}
\begin{eqnarray}
(p_1+p_2)^2 & = & s_{12}  \label{systemD1} , \\
(p_1+p_2 + p_3)^2 & = & s_{123}  \label{systemD2} .
\end{eqnarray}
\end{subequations}
By subtracting Eq.~(\ref{systemD2}) from Eq.~(\ref{systemD1}), we obtain 
an expression for $E_1$ that is a first order polynomial in $|\bs p_1 |$
and $|\bs p_2 |$. Inserting this expression into Eq.~(\ref{systemD1}) and into 
the equation defining the mass-shell condition for the particle 
of momentum $p_1$, we obtain a system of two
quadratic equations in $|\bs p_1 |$ and $|\bs p_2 |$
parametrized by the momentum $p_3$, by the invariants $s_{12}$,
$s_{123}$ and by the angles $\theta_1$, $\theta_2$, $\phi_1$ and $\phi_2$.
 This system can be solved analytically.
There are up to four solutions 
 for the modulus $|\bs p_1|$ and $|\bs p_2|$. Each solution that is physical ({\it i.e.}, such that $|\bs p_1|>0$ and $|\bs p_2|>0$)    
corresponds to a distinct  phase-space point
at which 
 the Jacobian in Eq.~(\ref{jacBLOCKD})
and the integrand must be evaluated.

\bibliography{data_base}
\end{document}